\documentclass[journal,onecolumn]{IEEEtran}

\usepackage[utf8]{inputenc}
\usepackage{amsmath}
\usepackage{amsfonts}
\usepackage{amssymb}
\usepackage{mathrsfs}
\usepackage{geometry}
\usepackage{paralist}
\usepackage{psfrag}
\usepackage{pstool}

\usepackage{enumitem}
\usepackage{mathtools}
\mathtoolsset{showonlyrefs=true,showmanualtags=true}

\usepackage{amsthm}
\usepackage{color}
\usepackage{hyperref}
\usepackage{xstring}
\usepackage{todonotes}
\usepackage{comment}
\usepackage{verbatim}
\usepackage{siunitx}
\usepackage{cite}

\newcommand{\h}[1][]{h_{#1}}

\renewcommand{\v}[1][]{v_{#1}}

\let\oldalpha\alpha
\renewcommand{\alpha}[1][]{\oldalpha_{#1}}
\let\oldsigma\sigma
\renewcommand{\sigma}[1][]{\oldsigma_{#1}}
\let\oldkappa\kappa
\renewcommand{\kappa}[1][]{\oldkappa_{#1}}
\let\oldbeta\beta
\renewcommand{\beta}[1][]{\oldbeta_{#1}}
\let\oldOmega\Omega
\renewcommand{\Omega}[1][]{\oldOmega_{#1}}
\let\oldomega\omega
\renewcommand{\omega}[1][]{\oldomega_{#1}}
\let\oldmu\mu
\renewcommand{\mu}[1][]{\oldmu_{#1}}
\let\oldnu\nu
\renewcommand{\nu}[1][]{\oldnu_{#1}}
\let\oldell\ell
\renewcommand{\ell}[1][]{\oldell_{#1}}
\let\oldeta\eta
\renewcommand{\eta}[1][]{\oldeta_{#1}}
\let\olddelta\delta
\renewcommand{\delta}[1][]{\olddelta_{#1}}
\let\oldlambda\lambda
\renewcommand{\lambda}[1][]{\oldlambda_{#1}}
\let\oldepsilon\epsilon
\renewcommand{\epsilon}[1][]{\oldepsilon_{#1}}
\let\oldrho\rho
\renewcommand{\rho}[1][]{\oldrho_{#1}}
\let\oldzeta\zeta
\renewcommand{\zeta}[1][]{\oldzeta_{#1}}
\let\oldxi\xi
\renewcommand{\xi}[1][]{\oldxi_{#1}}
\let\oldXi\Xi
\renewcommand{\Xi}[1][]{\oldXi_{#1}}
\let\oldphi\phi
\renewcommand{\phi}[1][]{\oldphi_{#1}}
\let\oldpsi\psi
\renewcommand{\psi}[1][]{\oldpsi_{#1}}
\let\oldpi\pi
\renewcommand{\pi}[1][]{\oldpi_{#1}}
\let\oldPi\Pi
\renewcommand{\Pi}[1][]{\oldPi_{#1}}
\let\oldchi\chi
\renewcommand{\chi}[1][]{\oldchi_{#1}}

\definecolor{myblue}{rgb}{0.21, 0.52, 0.95}

\renewcommand{\tilde}[1]{\widetilde{#1}}
\renewcommand{\hat}[1]{\widehat{#1}}

\renewcommand{\bar}[1]{\overline{#1}}
\newcommand{\pmtx}[1]{\begin{pmatrix}#1\end{pmatrix}}
\newcommand{\bmtx}[1]{\begin{bmatrix}#1\end{bmatrix}}
\newcommand{\tto}{\rightrightarrows}

\newtheorem{theorem}{Theorem}

\newtheorem{corollary}{Corollary}
\newtheorem{lemma}{Lemma}
\newtheorem{assumption}{Assumption}
\newtheorem*{assumption*}{Assumption}

\newtheorem{remark}{Remark}
\newtheorem{definition}{Definition}

\newcommand{\ceq}{:=}
\newcommand{\ceqinv}{=:}

\newcommand{\x}{\times}

\newcommand{\norm}[2][\mbox{}]{| #2|_{#1}}

\newcommand{\Kinf}{\mathcal{K}_\infty}

\newcommand{\signpower}[2]{\lfloor#1\rceil^{#2}}

\newcommand{\ball}{\mathbb{B}}
\newcommand{\Rnneg}{\mathbb{R}_{\geq 0}}

\newcommand{\reals}[1][]{\mathbb{R}^{#1}}
\newcommand{\naturals}{\mathbb{N}}

\newcommand{\minus}{\backslash}

\newcommand{\tp}{^\top}
\newcommand{\inv}{^{-1}}

\newcommand{\KL}{\mathcal{KL}}

\newcommand{\pl}{^+}
\DeclareMathOperator{\dom}{dom}

\DeclareMathOperator{\rge}{rge}

\DeclareMathOperator{\con}{\bar{con}}

\newcommand{\so}[1]{\mathfrak{so}(#1)}
\newcommand{\mf}[1]{\mathfrak{#1}}
\newcommand{\sk}{S}
\newcommand{\sphere}[1]{\mathbb{S}^{#1}}

\newcommand{\Amc}[1][]{\mathcal{A}_{#1}}

\newcommand{\Cmc}[1][]{\mathcal{C}_{#1}}
\newcommand{\Dmc}[1][]{\mathcal{D}_{#1}}
\newcommand{\Emc}[1][]{\mathcal{E}_{#1}}
\newcommand{\Fmc}[1][]{\mathcal{F}_{#1}}
\newcommand{\Gmc}[1][]{\mathcal{G}_{#1}}
\newcommand{\Hmc}[1][]{\mathcal{H}_{#1}}
\newcommand{\Imc}[1][]{\mathcal{I}_{#1}}
\newcommand{\Kmc}[1][]{\mathcal{K}_{#1}}

\newcommand{\Nmc}[1][]{\mathcal{N}_{#1}}

\newcommand{\Smc}[1][]{\mathcal{S}_{#1}}
\newcommand{\Tmc}[1][]{\mathcal{T}_{#1}}

\newcommand{\Umc}[1][]{\mathcal{U}_{#1}}
\newcommand{\Xmc}[1][]{\mathcal{X}_{#1}}
\newcommand{\Ymc}[1][]{\mathcal{Y}_{#1}}
\newcommand{\Zmc}[1][]{\mathcal{Z}_{#1}}
\usepackage{geometry}
\title{\LARGE \textbf{Model-Based Event-Triggered Implementation of Hybrid Controllers Using Finite-Time Convergent Observers}}
\author{Xuanzhi Zhu, Pedro Casau and Carlos Silvestre%
\thanks{X. Zhu is with the Institute for Systems and Robotics, Instituto Superior T\'{e}cnico, Universidade de Lisboa, Lisboa, Portugal. E-mail address: {\tt\small xuanzhi.zhu@tecnico.ulisboa.pt}. %
P. Casau is with the Instituto de Telecomunica\c{c}\~{o}es and Department of Electronics, Telecommunications and Informatics, University of Aveiro, Portugal. E-mail address: {\tt\small pcasau@ua.pt}. %
C. Silvestre is with the Department of Electrical and Computer Engineering, Faculty of Science and Technology, University of Macau, Macau, China, and is on leave from the Instituto Superior T\'{e}cnico, Universidade de Lisboa, Lisboa, Portugal. E-mail address: {\tt\small csilvestre@um.edu.mo}. %
This work was supported in part by the Macau Science and Technology Development Fund under Grant FDCT/0031/2020/AFJ, in part by the University of Macau, Macau, China, under Grant MYRG2020-00188-FST and Grant MYRG2022-00205-FST, in part by the Funda\c{c}\~{a}o para a Ci\^{e}ncia e a Tecnologia (FCT) through LARSyS - FCT Project UIDB/50009/2020, and in part by the Funda\c{c}\~{a}o para a Ci\^{e}ncia e a Tecnologia (FCT) through FCT Projects UIDB/50008/2020-UIDP/50008/2020 and UIBD/153759/2022.
}}

\date{}

\newcounter{verbose}
\newcommand{\xd}{x_d}
\newcommand{\xs}{x_s}
\newcommand{\xc}[1][]{x_{c#1}}

\newcommand{\xo}[1][]{
\IfEqCase{#1}{%
        {1}{\hat{x}_1}%
        {2}{\hat{x}_2}%
    }[x_o]%
}
\let\olddot\dot
\renewcommand{\dot}[1]{%
    \noexpandarg
    \IfEqCase{#1}{%
        {\xo[i]}{\olddot{\hat{x}}_{i}}%
        {\xo[1]}{\olddot{\hat{x}}_{1}}%
        {\xo[2]}{\olddot{\hat{x}}_{2}}%
        {\xc}{\olddot{x}_c}%
        {\xc[d]}{\olddot{x}_{cd}}%
        {\xd}{\olddot{x}_d}%
    }[\olddot{#1}]%
}
\newcommand{\settle}{T}

\begin{document}
\maketitle
\begin{abstract}
In this paper, we explore the conditions for asymptotic stability of the hybrid closed-loop system resulting from the interconnection of a nonlinear plant, an intelligent sensor that generates finite-time convergent estimates of the plant state, and a controller node that receives opportunistic samples from the sensor node when certain model-based event-triggering conditions are met. The proposed method is endowed with a degree of separation, in the sense that the controller design is independent of the sensor design. This is achieved under mild regularity conditions imposed on the hybrid closed-loop system and the existence of persistently flowing solutions. We demonstrate the versatility of the method by implementing it on: 1) a sampled-data controller for regulation of linear plants; 2) a synergistic controller for attitude stabilization of rigid bodies. The effectiveness of these novel controllers is demonstrated through numerical simulations.
\end{abstract}

\section{Introduction}
\subsection{Background}
Event-Triggered Control (ETC) has garnered significant interest in theoretical investigation due to its ability to reduce the data transmission rate compared to traditional periodic sampled-data control. 
Early works on ETC often assume full state feedback  (cf.~\cite{Tabuada2007}). However, this assumption may not hold for practical applications, prompting the development of output-based ETC. This paper focuses on generalizing one of the output-based ETC approaches, termed continuous-time observer-based ETC (cf.~\cite{Heemels2012}), to the hybrid systems domain and examines the conditions which guarantee closed-loop asymptotic stability, {robustness, and non-Zeno solutions}.

One important class of continuous-time observer-based ETC schemes adopts the fundamental idea rooted in the framework of model-based ETC (cf.~\cite{LUNZE2010}). The idea is to exploit knowledge of the plant to integrate a synthetic model at both the sensor and controller nodes, between which a communication channel is monitored by event-triggering conditions. The model calculates control actions when the channel is idle and updates its state when the channel is active.
One of the pioneer works in this direction~\cite{Lehmann2011} employs a fixed threshold to enable aperiodic transmission while ensuring ultimate boundedness of the state of a perturbed linear plant.
The subsequent work~\cite{Heemels2013} uses a relative threshold combined with periodic event detection to trigger transmission opportunistically and achieves global exponential stability and $\ell[2]$-gain performance for the linear plant.
The recent work~\cite{Zhu2024} examines how a fixed threshold with or without periodic event generation affects the stability, robustness, and the inter-transmission time for a linear plant.
{Due to utilization of information during idleness of communication channel, these continuous-time observer-based model-based ETC schemes show superiority in reducing transmissions over output-based ones~\cite{Wang2020,Fu2022,Yu2023,Koen2024} without an observer and over observer-based ones~\cite{ZHANG2014,Aranda2015,Zhao2021,Borri2024} without a synthetic model, as witnessed by its application to distributed systems in~\cite{Liu2020}, parameter-varying systems in~\cite{Lv2023}, etc.}

\subsection{Motivations}
For most of the relevant works on continuous-time observer-based model-based ETC, transmissions occur incessantly even in the absence of disturbance (cf.~\cite{Lehmann2011,Heemels2013,Liu2020,Lv2023}). This is due to the fact that state estimates converge asymptotically, hence the event-triggering condition may continue to issue transmissions. It is plausible that the number of transmissions is finite if the estimate converges to the plant state within a finite time. Evidence supporting this can be found in~\cite{Zhu2024}, where significant savings in communication efforts are observed. Hence, incorporating finite-time convergent sensor dynamics into the framework of continuous-time observer-based model-based ETC is the core motivation of this paper.

Hybrid systems encompass a combination of continuous-time and discrete-time dynamics that are described by differential and difference inclusions, respectively. Due to the sampling action by ETC, it becomes natural to model and analyze ETC systems under the framework of hybrid systems (cf.~\cite{Postoyan2015}). The major advantage of hybrid system models over other system models with impulsive dynamics is that, by meeting the hybrid basic conditions, {asymptotic stability of a compact set is not only uniform but also robust (cf.~\cite[Theorem~7.12]{Goebel2012}~and~\cite[Theorem~7.21]{Goebel2012}).} Nevertheless, there are few works on continuous-time observer-based model-based ETC that adopt the hybrid system approach. Most of the aforementioned works use a piecewise linear system or input-to-output stability approach to model and analyze the interconnection of continuous-time linear system components (cf.~\cite{Lehmann2011,Heemels2013,Liu2020,Lv2023}).
However, hybrid systems models are better suited to handle scenarios where some system components exhibit both continuous-time and discrete-time dynamics.
For instance, impulsive updates of the finite-time convergent observer in~\cite{zhu_cdc_2021} are treated within the hybrid system framework. The interconnection of a sampled-data controller with a hybrid observer in~\cite{zhu_acc_2022} is modeled and analyzed via hybrid tools.
Therefore, generalizing the framework of continuous-time observer-based model-based ETC to the hybrid systems domain is our second motivation of this paper.

The separation principle states that, if a stabilizing state feedback is fed with state estimates from a stabilizing observer, then resulting closed-loop system is stable (cf.~\cite{Teel2010}). This property is attractive from the design perspective since it allows for various combinations of controllers with observers. For linear systems, global separation can be achieved by decoupling the dynamics of the closed-loop system into independent controller and estimator components. For nonlinear systems, one can expect at most a local or semiglobal separation (cf.~\cite{Teel1994,Atassi1999}). A local separation principle for a class of hybrid systems in derived in~\cite{Teel2010}. However, it does not consider ETC while making  a conservative stability assumption on the observer.
In the context of continuous-time observer-based model-based ETC, the aforementioned works~\cite{Liu2020,Lv2023,zhu_acc_2022} fail in achieving a separation principle among system components. Notable exceptions are the works~\cite{Lehmann2011,Heemels2013,Zhu2024} on linear plants.
For nonlinear plants and hybrid system components in this paper, we do not insist on achieving a separation principle. Rather, we pursue a separation of the controller design from the sensor design, namely the global asymptotic stabilizing capability of the controller maintains when incorporating the sensor. This is our third motivation of this paper.

\subsection{{Challenges and approaches}}
We start by assuming that there exists a hybrid controller that asymptotically stabilizes a given compact set.  Furthermore, we assume that this controller has continuous access to the state of the plant. However, in most control applications, the state must be reconstructed from sensor measurements, which is a process that presents many challenges, such as 1) sensor selection, namely determining the set of outputs that are required to estimate the state of the plant; 2) sensor limitations, such as quantization, sampling frequency, among others. In this paper, we consider a new type of sensors which are able to estimate the state of the plant in finite-time and that transmits these estimates to the controller when certain state-dependent criteria are met. We aim to study under which conditions the properties of the nominal state feedback controller are not compromised by the intrusion of this class of sensors into the control loop. This pursuit encounters two major challenges, which we approach with the help of hybrid tools:
\begin{inparaenum}
	\item One challenge is to relate stability properties of the nominal set to those of the target set.
	On one hand, the hybrid time domain of a solution to the nominal system may differ from that of a solution to the closed-loop system, even if the sensor issues perfect state estimates.
	On the other hand, the distance of a vector to the target set may not be a bounded function of the distance of the respective vector components to the nominal set. The work~\cite{Teel2010} gets around these issues by making an stability assumption on the closed-loop system constrained by perfect recovery of state estimates, which goes against our pursuit.
	In this paper, we borrow the notions of $j$-reparametrization and Lipschitz set-valued mappings to bridge the gap between stability properties of the nominal set and those of the target set.
	\item The other challenge is to appropriately describe the stabilizing property of the hybrid sensor.
	The sensor has to accept inputs experiencing jumps (arising from event-triggered sampling) known as hybrid inputs (cf.~\cite{Bernard2020}), but there lacks a notion of stability/attractivity of sets for hybrid systems with hybrid inputs (cf.~\cite{Ricardo2021}).
	The method~\cite[Assumption~9]{Teel2010} takes the convex hull of the flow map with respect to the control input to get rid of the control input in the sensor dynamics. But this approach leads to a conservative stability assumption on the sensor, which can be hard to verify for practical examples. Another approach relies on the notion of complete uniform observability for a single-input single-output continuous-time plant (cf.~\cite[Assumption~2]{Marx2016}). However, this approach only admits specific system dynamics and hence it remains unknown how to adapt the approach to our case of general hybrid dynamics.
	In this paper, we propose an alternative approach that describes the stabilizing hybrid sensor in terms of set attractivity and invariance.
\end{inparaenum}

\subsection{Contributions}
Under the theory of hybrid systems, this paper proposes a novel continuous-time observer-based ETC framework that contributes on:
\begin{inparaenum}
	\item incorporating finite-time convergent sensor dynamics.
	{Few attempts as such have been made in the literature with notable exceptions in~\cite{Luo2022,Zhou2023}, but they do not follow the model-based ETC strategy};
	\item enabling the consideration of nonlinear plants, hybrid controllers, hybrid sensors, and model-based event-triggering conditions.
	{Such general system dynamics are rarely treated in the literature with notable exception in~\cite{zhu_acc_2022}, whereas most relevant works consider specific system dynamics (cf.~\cite{Lehmann2011,Heemels2013,Zhu2024,ZHANG2014,Aranda2015,Zhao2021,Borri2024,Liu2020,Lv2023})};
	\item separating the controller design from the sensor design.
	{This has barely been achieved in the literature on hybrid systems with notable exceptions in~\cite{Teel2010,Marx2016}, but their approaches have limitations as mentioned above that makes it hard to extend to our setting. The separation property obtained in this paper makes our approach versatile in selecting the hybrid controller for the nominal nonlinear plant.}
\end{inparaenum}
Using the proposed framework, we design novel model-based event-triggered controllers for regulation of linear plants as well as attitude stabilization of rigid bodies, which, to the best of our knowledge, have never been considered in the literature.
We reiterate that this paper does not develop particular structures for the hybrid controllers or sensors. Rather, it establishes rigorous conditions under which the target set is robustly asymptotically stable for the closed-loop system.

\section{Preliminaries \& Notation}\label{sec:notation}
Let $\naturals$, $\Rnneg$, $\reals$, $\reals[n]$, and $\reals[n\x m]$ denote the natural numbers, the nonnegative real numbers, the real numbers, the $n$-dimensional Euclidean space equipped with the inner product $\left\langle v,w\right\rangle\ceq v\tp w$ for each $v,w\in\reals[n]$, and the space of $n\x m$ matrices, respectively. The Euclidean norm is defined as $\norm{v}=\sqrt{\left\langle v,v\right\rangle}$ for each $v\in\reals[n]$.
Given nonempty $\Amc\subset\reals[n]$ and $\mu>0$, we define $\Amc+\mu\ball\ceq\{v\in\reals[n]: \norm[\Amc]{v}\leq \mu\}$, where $\norm[\Amc]{v}$ denotes the distance from $\v\in\reals[n]$ to $\Amc$ defined by  $\norm[\Amc]{v}\ceq\inf_{w\in\Amc}\norm{w-v}$.
The $n$-dimensional sphere is defined by $\sphere{n}\ceq\{v\in\reals[n+1]:\norm{v}= 1\}$.
Given $\Xmc\subset\reals[n+m]$ and $\Amc\subset\reals[m]$, define $\Pi(\Xmc|\Amc)\ceq\{v\in\reals[n]:(v,w)\in\Xmc\text{ for some }w\in\Amc\}$.
{Let $\Kinf$ denote the set of strictly increasing continuous functions $\alpha:[0,+\infty)\to[0,+\infty)$ satisfying $\alpha(0)=0$ and $\lim_{s\to+\infty}\alpha(s)=+\infty$.
Let $\KL$ denote the set of functions $\beta:[0,+\infty)^2\to[0,+\infty)$ that is nondecreasing in its first argument, nonincreasing in its second argument, $\lim_{r\searrow 0}\beta(r,s)=0$ for each $s\geq 0$, and $\lim_{s\to+\infty}\beta(r,s)=0$ for each $r\geq 0$.}
The tangent cone to a set $\Xmc\subset\reals[n]$ at a point $x\in\reals[n]$, denoted as $\Tmc[\Xmc](x)$, is the set of all vectors $w\in\reals[n]$ for which there exist sequences $x_i\in\Xmc$, $\tau_i>0$ with $x_i\to x$, $\tau_i\searrow 0$, and $w=\lim_{i\to+\infty}\frac{x_i-x}{\tau_i}$.
Given a set-valued mapping $\Fmc:\reals[n]\tto\reals[n]$, let $\dom\Fmc\ceq\{x\in\reals[n]:\Fmc(x)\neq\emptyset\}$ and $\rge\Fmc\ceq\{y\in\reals[n]:y\in\Fmc(x)\text{ for some } x\in\reals[n]\}$.
The continuous function
$\signpower{\cdot}{\beta}:\reals[n]\to\reals[n]$ for a given $\beta\in(0,1)$ is defined by 
$\signpower{v}{\beta}=(\mathrm{sign}(v_1)|v_1|^{\beta},\mathrm{sign}(v_2)|v_2|^{\beta},\cdots,\mathrm{sign}(v_n)|v_n|^{\beta})$
for each $v\in\reals[n]$.

The modeling and analysis hereafter, unless otherwise stated, follow the formalization of hybrid systems in~\cite{Goebel2012,Ricardo2021}. A hybrid system $\Hmc=(\Cmc,\Fmc,\Dmc,\Gmc)$ is defined by
\begin{equation}
\begin{aligned}
\dot{\xi}&\in \Fmc(\xi)& \xi&\in \Cmc\\
\xi\pl&\in \Gmc(\xi) & \xi&\in \Dmc
\end{aligned}
\end{equation}
where $\xi\in\reals[n]$ is the state, $\Cmc\subset\reals[n]$ is the flow set, $\Fmc:\reals[n]\tto\reals[n]$ is the flow map, $\Cmc\subset\reals[n]$ is the jump set, and $\Gmc:\reals[n]\tto\reals[n]$ is the jump map. A solution $\phi$ to $\Hmc$ is parametrized by the hybrid time $(t,j)$, where $t$ denotes the continuous time and $j$ denotes the number of jumps, and $\dom \phi \subset [0,+\infty)\times \mathbb{N}$ is a hybrid time domain: for each $(T,J)\in \dom \phi$, $\dom \phi\cap ([0,T]\times\{0,1,\dots J\})$ can be written in the form $\cup_{j=0}^{J-1}([t_j,t_{j+1}],j)$ for some finite sequence of times $0=t_0\leq t_1\leq t_2\leq \cdots \leq t_J$, where the $t_j$'s define the jump times.
A solution $\phi$ to $\Hmc$ is said to be \emph{maximal} if it cannot be extended by flowing nor jumping and \emph{complete} if its domain is unbounded.
Given $\Xmc\subset\reals[n]$, the set of maximal solutions to $\Hmc$ satisfying $\phi(0,0)\in \Xmc$ is denoted by $\Smc[\Hmc](\Xmc)$. The set of all maximal solutions to $\Hmc$ is denoted by $\Smc[\Hmc]$.

\begin{definition}[{\cite[Definitions~6.24-6.25]{Goebel2012}}, {\cite[Definition~3.1]{Ricardo2021}}, {\cite[Definition~3.1]{Li2019}}]\label{def:stabilities}
Given a hybrid system $\Hmc$, a nonempty $\Amc\subset\reals[n]$ is said to be:
\begin{itemize}[]
	\item stable for $\Hmc$ if for each $\varepsilon> 0$ there exists $\delta> 0$ such that each solution $\phi$ to $\Hmc$ with $\norm[\Amc]{\phi(0,0)}\leq\delta$ satisfies $\norm[\Amc]{\phi(t,j)}\leq\varepsilon$ for each $(t,j)\in\dom\phi$;
	\item uniformly stable for $\Hmc$ if there exists $\alpha\in\Kmc[\infty]$ such that each solution $\phi$ to $\Hmc$ satisfies $\norm[\Amc]{\phi(t,j)}\leq\alpha(\norm[\Amc]{\phi(0,0)})$ for each $(t,j)\in\dom\phi$;
	\item attractive for $\Hmc$ if there exists $\mu>0$ such that each $\phi\in\Smc[\Hmc](\Amc+\mu\ball)$ is complete and $\lim_{t+j\to+\infty}\norm[\Amc]{\phi(t,j)}=0$;
	\item uniformly attractive for $\Hmc$ from $\Xmc\subset\reals[n]$ if each solution $\phi\in\Smc[\Hmc](\Xmc)$ is complete, $(t,j)\mapsto\norm[\Amc]{\phi(t,j)}$ is bounded, and for each $\varepsilon>0$ there exists $\tau>0$ such that $\norm[\Amc]{\phi(t,j)}\leq\varepsilon$ for each $\phi\in\Smc[\Hmc](\Xmc)$ and $(t,j)\in\dom\phi$ with $t+j\geq\tau$;
	\item (uniformly) asymptotically stable for $\Hmc$ if it is both (uniformly) stable and (uniformly) attractive;
	\item finite-time attractive for $\Hmc$ from $\Xmc\subset\Nmc$ with $\Nmc\subset\reals[n]$ being an open neighborhood of $\Amc$ if there exists $T:\Nmc\to[0,+\infty)$, called the settling-time function, such that each solution $\phi\in\Smc[\Hmc](\Xmc)$ satisfies $\sup\{t+j:(t,j)\in\dom\phi\}\geq T(\phi(0,0))$ and $\lim_{(t,j)\in\dom\phi:t+j\nearrow T(\phi(0,0))}\norm[\Amc]{\phi(t, j)} =0$;
	\item strongly forward invariant for $\Hmc$ if each $\phi\in\Smc[\Hmc](\Amc)$ is complete and satisfies $\rge\phi\subset\Amc$.
\end{itemize}
\end{definition}
\begin{remark}
	If any of the quantities $\mu>0$, $\Xmc\subset\reals[n]$, and $\Nmc\subset\reals[n]$ above can be arbitrarily picked, then the respective definition admits a global version.
	By dropping the premise that each maximal solution is complete, the respective definition admits a ``pre-'' version.
\end{remark}

\begin{definition}[{\cite[Definition~3]{Bernard2020}}]\label{def:jreparametrization}
Given a hybrid arc $\phi$, a hybrid arc $\phi^r$ is a $j$-reparametrization of $\phi$ if there exists a function $\rho:\naturals\to\naturals$ such that: $\rho(0)=0$; $\rho(j+1)-\rho(j)\in\{0,1\}$ for each $j\in\naturals$;
and
$\phi^r(t,j)=\phi(t,\rho(j))$ for each $(t,j)\in\dom\phi^r.$
The hybrid arc $\phi^r$ is a full $j$-reparametrization of $\phi$ if
$\dom\phi = \bigcup_{(t,j)\in\dom\phi^r} (t,\rho(j)).$
\end{definition}

\begin{definition}[{\cite[Definition~9.26]{Rockafellar1998}}]\label{def:Lipschitz}
A set-valued mapping $\Pi:\reals[n]\tto\reals[m]$ is said to be Lipschitz continuous on $\Xmc\subset\reals[n]$ if it is nonempty-valued and closed-valued on $\Xmc$, and there exists $L\in\Rnneg$, called a Lipschitz constant, such that
$d(\Pi(x_1),\Pi(x_2))\leq L|x_1-x_2|$
for each $x_1,x_2\in\Xmc$, where $d:\Xmc^2\to\reals$ is defined by $d(\Xmc[1],\Xmc[2])=\sup_{x\in\reals[m]} |\norm[{\Xmc[1]}]{x}-\norm[{\Xmc[2]}]{x}|$ for each nonempty $\Xmc[1],\Xmc[2]\subset\Xmc$.
\end{definition}
\begin{remark}\label{rem:Lip}
The notion above agrees with Lipschitz continuity of a function when a Lipschitz continuous set-valued mapping is single-valued.
\end{remark}

\begin{definition}[Inspired by~{\cite[Definition~3]{Postoyan2015}}~and~{\cite[Proposition~3.27]{Goebel2012}}]\label{def:finite jumps}
The solutions to a hybrid system $\Hmc$ are said to possess a uniform semiglobal persistent flow time from $\Xmc\subset\reals[n]$ if, for each $r\in\Rnneg$ there exist $\tau>0$ and $N\in\naturals$ such that, for each solution $\psi$ to $\Hmc$ with $\psi(0,0)\in \Xmc\cap r\ball$, $j\leq \frac{t}{\tau}+N$ for each $(t,j)\in\dom\psi$.
\end{definition}

We introduce a technical lemma that separates iterated infima involving a set-valued relation between the arguments.

\begin{lemma}\label{lem:iterated infima set-valued}
Given sets $\Xmc\subset\reals[n]$ and $\Ymc\subset\reals[m]$, a function ${g}:\Xmc\x\Ymc\to\reals$, and a set-valued mapping $\Pi:\reals[n]\tto\reals[m]$, then $\inf_{x\in\Xmc,y\in\Pi(x)} {g}(x,y) = \inf_{x\in\Xmc}\inf_{y\in\Pi(x)} {g}(x,y)$.
\end{lemma}
\begin{IEEEproof}
Let $\Zmc\ceq\{g(x,y):x\in\Xmc,y\in\Pi(x)\}$, $g_x:\Xmc\to\reals$ be defined by $g_x(x)=\inf\{g(x,y):y\in\Pi(x)\}$ for each $x\in\Xmc$. Since $g(x,y)\geq g_x(x)$ for each $x\in\Xmc$ and each $y\in\Pi(x)$, it follows that $\inf\{g(x,y):x\in\Xmc,y\in\Pi(x)\}\geq\inf\{g_x(x):x\in\Xmc,y\in\Pi(x)\}$, implying that $\inf\Zmc\geq\inf\{g_x(x):x\in\Xmc\}$.
Next we prove that $\inf\Zmc\leq\inf\{g_x(x):x\in\Xmc\}$. Suppose for sake of contradiction that $\inf\Zmc>\inf\{g_x(x):x\in\Xmc\}$, then there exists $x'\in\Xmc$ such that $\inf\Zmc>g_x(x')=\inf\{g(x',y):y\in\Pi(x')\}$, which in turn implies the existence of $y'\in\Pi(x')$ such that $\inf\Zmc>g(x',y')$. Hence $\inf\Zmc$ is not a lower bound for $\Zmc$, which is a contradiction. This implies that $\inf\Zmc\leq\inf\{g_x(x):x\in\Xmc\}$ and consequently $\inf\Zmc=\inf\{g_x(x):x\in\Xmc\}$.
\end{IEEEproof}

\section{Problem Setup}\label{sec:mbetc}
Consider a nominal plant defined by
\begin{equation}\label{eqn:plant0}
\dot{x}=f(x,u)
\end{equation}
where $x\in\reals[n_x]$ and $u\in\reals[n_u]$ denote the nominal plant state and the control input, respectively. 

Suppose that there exists a hybrid controller $\Hmc[c]\ceq(\Cmc[c],\Fmc[c],\Dmc[c],\Gmc[c])$, defined by
\begin{equation}\label{eqn:Hmcc}
\begin{aligned}
\dot{\xc}&\in\Fmc[c](\xc,x)& (\xc,x)&\in \Cmc[c]\\
\xc\pl&\in \Gmc[c](\xc,x) & (\xc,x)&\in \Dmc[c]
\end{aligned}
\end{equation}
where $\xc\in\reals[n_c]$ denotes the controller state.
The hybrid controller outputs $u$ through the feedback law $\kappa:\reals[n_c]\x\reals[n_x]\to\reals[n_u]$
\begin{equation}
    \label{eqn:u}
    u\ceq\kappa(\xc,x)
\end{equation}

The interconnection between~\eqref{eqn:plant0} and~\eqref{eqn:Hmcc} through~\eqref{eqn:u} is represented by the nominal hybrid system $\Hmc[0]\ceq(\Cmc[0],\Fmc[0],\Dmc[0],\Gmc[0])$, defined by
%
\begin{equation}
\label{eqn:Hmc0}
\begin{aligned}
\Fmc[0](x,\xc)&\ceq(f(x,\kappa(\xc,x)),\Fmc[c](\xc,x)) & &\forall (x,\xc)\in\Cmc[0]\\
\Gmc[0](x,\xc)&\ceq(x,\Gmc[c](\xc,x)) & &\forall (x,\xc)\in\Dmc[0]
\end{aligned}
\end{equation}
where $\Cmc[0]\ceq\{(x,\xc)\in\reals[n_x+n_c]:(\xc,x)\in\Cmc[c]\}$ and $\Dmc[0]\ceq\{(x,\xc)\in\reals[n_x+n_c]:(\xc,x)\in\Dmc[c]\}$.

The next assumption endows the controller $\Hmc[c]$ with the capability to stabilize some compact set, {which is a standard assumption to make for the nominal hybrid system, see~\cite[Assumption~2.3]{Atassi1999},~\cite[Assumption~3]{Teel2010}, or~\cite[Assumption~1]{Marx2016}.}

\begin{assumption}
\label{ass:Amc0}
There exists a nonempty compact set $\Amc[0]\subset\reals[n_x+n_c]$ that is globally asymptotically stable for $\Hmc[0]$ in~\eqref{eqn:Hmc0}.
\end{assumption}
{\begin{remark}\label{rem:Amc0}
The assumption above is an evolution from our earlier work~\cite{zhu_acc_2022} and makes a major improvement in terms of simplicity in the controller node design, since one has to take the sensor node into account for controller design as in~\cite[Assumption~3]{zhu_acc_2022}.
\end{remark}}

Now suppose that the nominal plant state in~\eqref{eqn:plant0} is not available for feedback. Rather, there is an output equation added to~\eqref{eqn:plant0}, leading to the plant $\Hmc[p]$ defined by
\begin{equation}\label{eqn:plant}
\begin{aligned}
\dot{x}&=f(x,u)\\
y&=h(x)
\end{aligned}
\end{equation}
where $x\in\reals[n_x]$, $u\in\reals[n_u]$, and $y\in\reals[n_y]$ denote the plant state, the control input, and the plant output, respectively. Due to the lack of plant state information, the hybrid state feedback in~\eqref{eqn:Hmcc} and~\eqref{eqn:u} can no longer be used to stabilize the plant. To fix this, we consider the following form of hybrid observer $\Hmc[o]\ceq(\Cmc[o],\Fmc[o],\Dmc[o],\Gmc[o])$ that uses the plant output to recover the plant state, feeding finite-time convergent estimates to the hybrid controller.
\begin{equation}\label{eqn:Hmco}
\begin{aligned}
\dot{x}_o&\in\Fmc[o](\xo,y,u) & (\xo,y,u)&\in \Cmc[o]\\
\xo\pl&\in \Gmc[o](\xo,y,u) & (\xo,y,u)&\in \Dmc[o]
\end{aligned}
\end{equation}
where $\xo\in\reals[n_o]$ denotes the observer state.
The observer outputs $\hat{x}\in\reals[n_{\hat{x}}]$ through the law $\h[o]:\reals[n_o]\x\reals[n_y]\to\reals[n_{\hat{x}}]$ defined by $\hat{x}\ceq\h[o](\xo,y)$.

Furthermore, we assume that there is a communication channel between the hybrid observer and the hybrid controller. Taking this into account, we follow the model-based ETC approach by further introducing a synthetic model of the plant, $\Hmc[s]\ceq(\Cmc[s],f,\Dmc[s],\Gmc[s])$, defined by
\begin{equation}\label{eqn:Hmcs}
\begin{aligned}
\dot{x}_s&= f(\xs,u) & (\xs,u,\hat{x})&\in \Cmc[s]\\
\xs\pl&\in\Gmc[s](\hat{x}) & (\xs,u,\hat{x})&\in \Dmc[s]
\end{aligned}
\end{equation}
where $\xs\in\reals[n_x]$ denotes the synthetic state.
The flow dynamics of $\Hmc[s]$ replicates the behavior of the nominal plant. In this way, replacing $x$ with $\xs$ in both the hybrid controller $\Hmc[c]$ in~\eqref{eqn:Hmcc} and the feedback law~\eqref{eqn:u} does not alter the behavior of the closed-loop system provided that $\xs$ equals $x$ for all time.
The jump dynamics of $\Hmc[s]$ specify how and when information from the observer is transmitted to the controller.

We consider that the sensor node consists of the hybrid observer, a copy of the synthetic model, and a copy of the hybrid controller while the controller node consists of a copy of the synthetic model and and a copy of the hybrid controller.

\begin{figure}[ht!]
	\centering
	\includegraphics[width=0.7\textwidth]{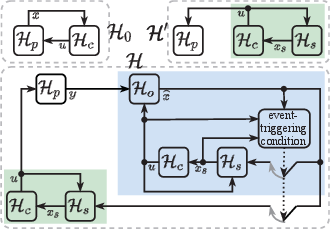}
	\caption{{A graphical illustration of the nominal system $\Hmc[0]$ in~\eqref{eqn:Hmc0} (upper left), the closed-loop system $\Hmc$ in~\eqref{eqn:closedloop} (bottom), and the reduced system $\Hmc'$ after finite-time convergence of the sensor dynamics (upper right). Smaller shaded rectangle: controller node. Larger shaded rectangle: sensor node. In $\Hmc$, the replicas of $\Hmc[c]$ are synchronized and the replicas of $\Hmc[s]$ are synchronized.}}
	\label{fig:diagram}
\end{figure}

By replacing $x$ with $\xs$ in both the hybrid controller $\Hmc[c]$ in~\eqref{eqn:Hmcc} and the feedback law~\eqref{eqn:u}, we interconnect the plant $\Hmc[p]$ in~\eqref{eqn:plant}, the hybrid controller $\Hmc[c]$ in~\eqref{eqn:Hmcc}, the hybrid observer $\Hmc[o]$ in~\eqref{eqn:Hmco}, and the synthetic model $\Hmc[s]$ in~\eqref{eqn:Hmcs}.
Mathematically, the interconnection results in a closed-loop system $\Hmc\ceq(\Cmc,\Fmc,\Dmc,\Gmc)$ defined by
\begin{equation}\label{eqn:closedloop}
	\begin{aligned}
	&\Fmc(\xi)\ceq\pmtx{f(x,\kappa(\xc,\xs))\\
		\Fmc[c](\xc,\xs)\\
		f(\xs,\kappa(\xc,\xs))\\
		\Fmc[o](\xo,h(x),\kappa(\xc,\xs))} & &\forall \xi\in\Cmc\\
	&\Gmc(\xi)\ceq\tilde{\Gmc}_c(\xi)\cup \tilde{\Gmc}_{so}(\xi)& &\forall \xi\in\Dmc
	\end{aligned}
\end{equation}
where the flow and jump sets are defined by 
\begin{align*}
 \Cmc&\ceq\{\xi\in\Xi: (\xc,\xs)\in\Cmc[c],(\xs,\kappa(\xc,\xs),\h[o](\xo,h(x)))\in\Cmc[s],(\xo,h(x),\kappa(\xc,\xs))\in\Cmc[o]\}\\
 \Dmc&\ceq\{\xi\in\Xi: (\xc,\xs)\in\Dmc[c]\text{ or }\xi\in\Dmc[so]\}
\end{align*}
respectively, and $\tilde{\Gmc}_c,\tilde{\Gmc}_{so}:\Xi\tto\Xi$ are defined by 
 \begin{align*}
     \tilde{\Gmc}_c(\xi)&\ceq(x,\Gmc[c](\xc,\xs),\xs,\xo)\\
     \tilde{\Gmc}_{so}(\xi)&\ceq(x,\xc,\Gmc[so](\xi))
 \end{align*}
for each $\xi\in\Xi$, where $\Dmc[so]\subset\Xi$ and $\Gmc[so]:\Xi\to\reals[n_x+n_o]$ with $\dom\Gmc[so]=\Dmc[so]$ are designed based on $\Dmc[o]$ and $\Dmc[s]$.
{Graphically, we illustrate the constructions above in Fig.~\ref{fig:diagram}. Specifically, for the given nominal system $\Hmc[0]$, we implement model-based ETC on $\Hmc[0]$. The resulting closed-loop system $\Hmc$, after finite-time convergence of the sensor dynamics, can be identified as a reduced system $\Hmc'$ that operates in open-loop only with the controller node.}
The following assumption enforces some regularity conditions on $\Hmc$ that are pivotal in proving existence of complete solutions to and robustness of hybrid systems, {which is a standard practice in the ETC literature, see~\cite{Postoyan2015,Ricardo2021,Zhu2024}.}

\begin{assumption}[{\cite[Assumption~6.5,~Proposition~6.10]{Goebel2012}}]\label{ass:hbc}
The hybrid system $\Hmc$ in~\eqref{eqn:closedloop} satisfies the hybrid basic conditions given below.
\begin{itemize}[leftmargin=*]
    \item Both $\Cmc$ and $\Dmc$ are closed relative to $\reals[n]$;
    \item $\Fmc$ is outer semicontinuous and locally bounded relative to $\Cmc\subset\dom\Fmc$ and $\Fmc$ is convex-valued on $\Cmc$;
    \item $\Gmc$ is outer semicontinuous and locally bounded relative to $\Dmc\subset\dom\Gmc$.
\end{itemize}
Moreover, the following viability conditions hold for $\Hmc$.
\begin{itemize}[leftmargin=*]
    \item For each $\xi[0]\in\Cmc\minus\Dmc$ there exists a neighborhood $\Umc$ of $\xi[0]$ such that $\Fmc(\xi)\cap\Tmc[\Cmc](\xi)\neq\emptyset$ for each $\xi\in\Cmc\cap\Umc$;
    \item either $\Cmc$ is compact, or $\Fmc$ has linear growth on $\Cmc$, or $\Fmc$ is single-valued and locally Lipschitz on $\Cmc$;
    \item $\Gmc(\Dmc)\subset\Cmc\cup\Dmc$.
\end{itemize}
\end{assumption}

{We define the set where the sensor achieves finite-time convergence as follows.}
\begin{align}
    \label{eqn:Amc_s}
    \Amc[s]\ceq\{\xi\in\Xi:x=\xs,\xo\in\Pi(x)\}
\end{align}
where $\Pi:\reals[n_x]\tto\reals[n_o]$ is outer semicontinuous, locally bounded, Lipschitz continuous on $\reals[n_x]$ (cf.~Definition~\ref{def:Lipschitz}), and nonempty-valued on $\reals[n_x]$.
The introduction of $\Pi$ allows one to: add extra degrees of freedom; change coordinates; do a partial state estimate; and handle set-valued dynamics, when designing the observer.

{Our goal is to derive conditions on $\Hmc$ such that the set
	\begin{align}
	\label{eqn:Amc}
	\Amc\ceq\{\xi\in\Xi:(x,\xc)\in\Amc[0],\xi\in\Amc[s]\}
	\end{align}
	is robustly asymptotically stable. Note that, in this way, we preserve the stabilizing capability of the original state-feedback controller in~\ref{eqn:Hmcc} in the presence of the proposed event-triggered sensor.
}

\section{Main Results}\label{sec:stability}
This section presents the conditions for asymptotic stability of $\Amc$ for $\Hmc$, robustness of $\Hmc$ to small perturbations, and absence of Zeno solutions.
Specifically, we show that if the sensor dynamics are finite-time attractive with a bounded settling-time function, then it is possible to asymptotically stabilize $\Amc$ for the closed-loop system, provided that, after finite-time stabilization of the sensor dynamics, solutions stay within $\Amc[s]$ and the distance to $\Amc$ can be upper bounded by the distance to $\Amc[0]$. 
We proceed by first studying the sensor dynamics, which is reflected in the following assumption that asks for the properties $\Amc[s]$ should have.

\begin{assumption}\label{ass:Amcs}
There exists $\mu>0$ and $\ell>\sup_{\xi\in\Amc+\mu\ball}|\xi|$ such that $\Amc[s]$ in~\eqref{eqn:Amc_s} is finite-time attractive for $\Hmc$ in~\eqref{eqn:closedloop} from $(\Amc[s]+\mu\ball)\cap \ell\ball$ with settling-time function $\settle:\Xi\to[0,+\infty)$ satisfying
\begin{equation}
    \label{eqn:finite_settle}
  \sup\settle((\Amc[s]+\mu\ball)\cap \ell\ball)<+\infty
\end{equation}
and $\Amc[s]$ is strongly forward pre-invariant for $\Hmc$.
\end{assumption}
{\begin{remark}
    Compared to Assumption~\ref{ass:Amc0} which acts solely on the controller node by studying the nominal system $\Hmc[0]$, the assumption above on $\Hmc$ involves both the controller node and the sensor node. It seems then that we do not separate the design for these nodes but we gain some degree of separation because the controller node is designed independently from the sensor node.
\end{remark}}

In order to associate $\Amc[s]$ in~\eqref{eqn:Amc_s} in the assumption above with $\Amc[0]$ in Assumption~\ref{ass:Amc0}, we next study the ``tail'' of a solution in $\Smc[\Hmc]((\Amc[s]+\mu\ball)\cap\ell\ball)$. This is motivated by the fact that, after finite-time convergence to $\Amc[s]$, $x$ matching $\xs$ implies $(\xs,\xc)\in\Amc[0]$.

\begin{lemma}\label{lem:phi}
If Assumptions~\ref{ass:hbc} and~\ref{ass:Amcs} hold, then for each $\phi\in\Smc[\Hmc]((\Amc[s]+\mu\ball)\cap\ell\ball)$ there exists $(t^\star,j^\star)\in\dom\phi$ with $t^\star+j^\star\geq\settle(\phi(0,0))$ such that the hybrid arc $\psi$ defined by
\begin{align}\label{eqn:phi}
    (t,j)\mapsto \psi(t,j)\ceq\phi(t+t^\star,j+j^\star)
\end{align}
for each $(t,j)\in\dom\psi\ceq\{(t,j)\in\Rnneg\times \naturals: (t+t^\star,j+j^\star)\in\dom\phi\}$ is a complete solution to $\Hmc$. Moreover, $\psi\in\Smc[\Hmc](\Amc[s])$.
\end{lemma}
\begin{IEEEproof}
Due to Assumption~\ref{ass:hbc},
we are in the case (a) of~\cite[Proposition~6.10]{Goebel2012} where each solution $\phi\in\Smc[\Hmc]((\Amc[s]+\mu\ball)\cap\ell\ball)$ is complete, implying that there exists $(t^\star,j^\star)\in\dom\phi$ satisfying $t^\star+j^\star\geq\settle(\phi(0,0))$, hence $\phi(t^\star,j^\star)\in\Amc[s]$ by Definition~\ref{def:stabilities} on finite-time attractivity.
For each $j\in\naturals$ such that $\Imc[\psi]^j=\{t:(t,j)\in\dom\psi\}$, it follows that $\Imc[\psi]^j=\{t:(t+t^\star,j+j^\star)\in\dom\phi\}=\{t-t^\star:(t,j+j^\star)\in\dom\phi\}$
has nonempty interior if and only if $\Imc[\phi]^{j+j^\star}=\{t:(t,j+j^\star)\in\dom\phi\}$ does, since $\Imc[\psi]^j+t^\star=\Imc[\phi]^{j+j^\star}$. Now for each $j\in\naturals$ such that $\Imc[\psi]^j$ has nonempty interior, it holds
\begin{align*}
    \psi(t,j)&=\phi(t+t^\star,j+j^\star)\in\Cmc & & \forall t\in\text{int }\Imc[\psi]^j\\
    \dot{\psi}(t,j)&=\dot{\phi}(t+t^\star,j+j^\star)\in\Fmc(\psi(t,j)) & &
\end{align*}
for almost all $t\in\Imc[\psi]^j$.
For each $(t,j)\in\dom\psi$ such that $(t,j+1)\in\dom\psi$, $(t+t^\star,j+j^\star)\in\dom\phi$ with $(t+t^\star,j+j^\star+1)\in\dom\phi$, hence
\begin{align*}
    \psi(t,j)&=\phi(t+t^\star,j+j^\star)\in\Dmc\\
    \psi(t,j+1)&=\phi(t+t^\star,j+j^\star+1)\in\Gmc(\psi(t,j))
\end{align*}
which identifies $\psi$ as a solution to $\Hmc$, where completeness of $\psi$ follows from completeness of $\phi$ and from the fact that the left-hand side of~\eqref{eqn:finite_settle} is finite. The claim $\psi\in\Smc[\Hmc](\Amc[s])$ comes from the fact that $\psi(0,0)=\phi(t^\star,j^\star)\in\Amc[s]$.
\end{IEEEproof}

To put it simply, the ``tail'' of a solution in $\Smc[\Hmc]((\Amc[s]+\mu\ball)\cap\ell\ball)$ is a solution in $\Smc[\Hmc](\Amc[s])$, which facilitates the presentation of the following lemma.

\begin{lemma}\label{lem:phi0}
If Assumptions~\ref{ass:hbc} and~\ref{ass:Amcs} hold, then for each $\phi\in\Smc[\Hmc](\Amc[s])$ with $\phi(t,j)=(x,\xc,\xs,\xo)(t,j)$ for each $(t,j)\in\dom\phi$, there exists a solution $\phi[0]$ to $\Hmc[0]$ such that
\begin{align}\label{eqn:phi0}
    \phi(t,j)&=(\phi[0]^r,\xs,\xo)(t,j) & &\forall (t,j)\in\dom\phi
\end{align}
where $\phi[0]^r$ is a full $j$-reparametrization of $\phi[0]$.
\end{lemma}
\begin{IEEEproof}
Since $\Amc[s]$ is strongly forward pre-invariant for $\Hmc$ by Assumption~\ref{ass:Amcs}, we have each solution $\phi\in\Smc[\Hmc](\Amc[s])$ satisfies $x(t,j)=\xs(t,j)$ for each $(t,j)\in\dom\phi$. Hence
\begin{align*}
    \dfrac{d}{dt}x(t,j)&=f(x(t,j),\kappa(\xc(t,j),x(t,j)))\\
    \dfrac{d}{dt}\xc(t,j)&\in\Fmc[c](\xc(t,j),x(t,j))
\end{align*}
for each $(t,j)\in\dom\phi$ where these derivatives are defined. Moreover, one of the following holds
\begin{align}\label{eqn:(x,x_c) by G_s}
(x,\xc)(t,j+1)&=(x,\xc)(t,j)\\
(x,\xc)(t,j+1)&\in (x(t,j),\Gmc[c](\xc(t,j),x(t,j)))
\end{align}
when $\phi(t,j)\in\Dmc$ such that $\phi(t,j+1)\in\Gmc(\phi(t,j))$.
Consequently, for each solution $\phi\in\Smc[\Hmc](\Amc[s])$ with $\phi(t,j)=(x,\xc,\xs,\xo)(t,j)$ for each $(t,j)\in\dom\phi$, the only difference between the components $(x,\xc)$ of $\phi$ and a solution $\phi[0]$ to $\Hmc[0]$ satisfying $\phi[0](0,0)=(x,\xc)(0,0)$ is that $(t,j)\mapsto (x,\xc)(t,j)$ may have additional jumps which keep the value of $(x,\xc)$ unchanged as in~\eqref{eqn:(x,x_c) by G_s}. Therefore, we define $\rho:\naturals\to\naturals$ such that: $\rho(0)=0$,
$\rho(j+1)=\rho(j)$ if there exists $t\in[0,+\infty)$ such that $\phi(t,j)\in\Dmc[so]$ and $\phi(t,j+1)\in\tilde{\Gmc}_{so}(\phi(t,j))$,
and $\rho(j+1)=\rho(j)+1$ if there exists $t\in[0,+\infty)$ such that $(\xc,\xs)(t,j)\in\Dmc[c]$ and $\phi(t,j+1)\in\tilde{\Gmc}_c(\phi(t,j))$. It follows that $(x,\xc)(t,j)=\phi[0](t,\rho(j))$ for each $(t,j)\in\dom\phi$. Letting $\phi[0]^r:\dom\phi\to\reals[n_x+n_c]$ be defined by $(t,j)\mapsto\phi[0]^r(t,j)\ceq\phi[0](t,\rho(j))$ completes the proof.
\end{IEEEproof}

In other words, we are able to match the ``tail'' of a solution in $\Smc[\Hmc]((\Amc[s]+\mu\ball)\cap\ell\ball)$ with a solution to the nominal system $\Hmc[0]$, which relates $\Amc[s]$ in~\eqref{eqn:Amc_s} and $\Amc[0]$ in Assumption~\ref{ass:Amc0} to each other.
To study stability properties of $\Amc$ in~\eqref{eqn:Amc}, we should also investigate the connection between $\Amc$ and $\Amc[0]$. The following lemma ensures that the geometry of the sets $\Amc$ and $\Amc[0]$ does not compromise the stability properties that we present in the sequel.

\begin{lemma}\label{lem:norm}
If Assumptions~\ref{ass:Amc0}-\ref{ass:Amcs} hold, then there exists $\alpha\in\Kinf$ such that for each $\phi\in\Smc[\Hmc](\Amc[s])$ satisfying~\eqref{eqn:phi0}, it holds 
\begin{equation}\label{eqn:norm}
    \norm[\Amc]{\phi(t,j)}\leq \alpha(\norm[{\Amc[0]}]{\phi[0]^r(t,j)})
\end{equation}
for each $(t,j)\in\dom\phi$, where $\phi[0]^r$ is the $j$-reparametrization of $\phi[0]$ from Lemma~\ref{lem:phi0}.
\end{lemma}
\begin{IEEEproof}
Recall from the proof of Lemma~\ref{lem:phi0} that each $\phi\in\Smc[\Hmc](\Amc[s])$ with $\phi(t,j)=(x,\xc,\xs,\xo)(t,j)$ for each $(t,j)\in\dom\phi$ satisfies $x(t,j)=\xs(t,j)$ for each $(t,j)\in\dom\phi$.
Fix $\theta\ceq(\theta_x,\theta_c,\theta_x,\theta_o)\in\Amc[s]$ with $\theta_o\in\Pi(\theta_x)$ and let $\theta_0^r\ceq(\theta_x,\theta_c)$.
For arbitrary $\gamma\in\Amc$, let $\gamma\ceq(\gamma_x,\gamma_c,\gamma_x,\gamma_o)$ with $\gamma_o\in\Pi(\gamma_x)$ and let $\gamma_0^r\ceq(\gamma_x,\gamma_c)\in\Amc[0]$. Then
{\allowdisplaybreaks\begin{align}\label{eqn:norm deduction}
\norm[\Amc]{\theta}&=\inf_{\gamma\in\Amc}\norm{\gamma-\theta}\\
&=\inf_{\gamma\in\Amc} \sqrt{2|\theta_x-\gamma_x|^2+|\theta_c-\gamma_c|^2+|\theta_o-\gamma_o|^2}\\
&\leq\inf_{\substack{(\gamma_x,\gamma_c)\in\Amc[0],\\ \gamma_o\in\Pi(\gamma_x)}} \sqrt{2|(\theta_x-\gamma_x,\theta_c-\gamma_c)|^2+|\theta_o-\gamma_o|^2}\\
&=\inf_{\substack{\gamma_0^r\in\Amc[0],\\ \gamma_o\in\Pi(\gamma_x)}} \sqrt{2|\theta_0^r-\gamma_0^r|^2+|\theta_o-\gamma_o|^2}
\end{align}}
Consider the set $\Gamma\ceq\bigcup_{\gamma_0^r\in\Amc[0]}\Pi(M\gamma_0^r)$
with $M\ceq\bmtx{I&0}$ such that $\gamma_x=M\gamma_0^r$.
Define $g:\Amc[0]\x\Gamma\to\reals$ by
$$g(\gamma_0^r,\gamma_o)=\sqrt{2|\theta_0^r-\gamma_0^r|^2+|\theta_o-\gamma_o|^2}$$
for each $(\gamma_0^r,\gamma_o)\in\Amc[0]\x\Gamma$.
Since $\Pi$ is Lipschitz continuous, there exists $L\in\Rnneg$ which is a Lipschitz constant of $\Pi$. We derive from the last equality of~\eqref{eqn:norm deduction} that
\begin{equation}\label{eqn:norm deduction 2}
\begin{aligned}
\norm[\Amc]{\theta}&\leq\inf_{\gamma_0^r\in\Amc[0]}\inf_{\gamma_o\in\Pi'(\gamma_0^r)} \sqrt{2|\theta_0^r-\gamma_0^r|^2+|\theta_o-\gamma_o|^2}\\
&\leq\inf_{\gamma_0^r\in\Amc[0]} \sqrt{2|\theta_0^r-\gamma_0^r|^2+L^2|\theta_x-\gamma_x|^2}\\
&\leq\sqrt{2+L^2}\inf_{\gamma_0^r\in\Amc[0]} |\theta_0^r-\gamma_0^r|\\
&=\sqrt{2+L^2} \norm[{\Amc[0]}]{\theta_0^r}
\end{aligned}
\end{equation}
where the first inequality holds due to Lemma~\ref{lem:iterated infima set-valued}, and
the second inequality is proved by recalling that $\theta_o\in\Pi(\theta_x)$ and applying Definition~\ref{def:Lipschitz}
with $x_1=\theta_x$ and $x_2=\gamma_x$. Specifically, we have
\begin{equation}\notag
\begin{aligned}
&\inf_{\gamma_o\in\Pi'(\gamma_0^r)}|\theta_o-\gamma_o|=\inf_{\gamma_o\in\Pi(\gamma_x)}|\theta_o-\gamma_o|\\
&=\norm[\Pi(\gamma_x)]{\theta_o}\\
&=|\norm[\Pi(\theta_x)]{\theta_o}-\norm[\Pi(\gamma_x)]{\theta_o}|\\
&\leq \sup_{\theta_o\in\reals[n_o]}|\norm[\Pi(\theta_x)]{\theta_o}-\norm[\Pi(\gamma_x)]{\theta_o}|\\
&=d(\Pi(\theta_x),\Pi(\gamma_x))\\
&\leq L|\theta_x-\gamma_x|
\end{aligned}
\end{equation}

Setting $s\mapsto\alpha(s)\ceq\sqrt{2+L^2}s$ for each $s\in\Rnneg$ concludes the proof.
\end{IEEEproof}

Note that Assumption~\ref{ass:Amc0} ignores the jumps that maintains the value of $(x,\xc)$. In order for these jumps to not impede asymptotic stability of $\Amc$ for $\Hmc$, we make the last assumption as follows.


\begin{assumption}\label{ass:finite jumps}
The solutions to $\Hmc$ in~\eqref{eqn:closedloop} possess a uniform semiglobal persistent flow time from $\Amc[s]$ in~\eqref{eqn:Amc_s}.
\end{assumption}
{\begin{remark} 
The above assumption essentially says that the number of jumps should be small enough, while this number is uniform with respect to all solutions starting from each compact set within $\Amc[s]$.
The above assumption is rather standard in the literature that seeks a separation between the controller design and the sensor design under the framework of hybrid systems. For instance, \cite[Assumption~4]{Teel2010} asks for the absence of complete discrete solutions for both the controller dynamics and the sensor dynamics;
\cite[Equation~(2c)]{Fichera2013a} inhibits any jumps within a continuous-time interval of positive length; and~\cite[Assumption~1]{Marx2016} requires that any controller jumps must be separated in continuous time by a positive amount.
\end{remark}}
For this paper, Assumption~\ref{ass:finite jumps} turns out to be essential in proving
uniformity of attractivity.
Based on the aforementioned discussions, we now present the main result of this paper as follows.

\begin{theorem}\label{thm:stability}
If Assumptions~\ref{ass:Amc0}-\ref{ass:finite jumps} hold, then $\Amc$ in~\eqref{eqn:Amc} is uniformly locally asymptotically stable for $\Hmc$ in~\eqref{eqn:closedloop}, with a basin of attraction containing $\Amc+\mu\ball$, where $\mu$ comes from Assumption~\ref{ass:Amcs}.
\end{theorem}
\begin{IEEEproof}
Note that $\Amc$ is compact by construction.
First, we prove that $\Amc$ is strongly forward pre-invariant for $\Hmc$. Pick arbitrary $\phi\in\Smc[\Hmc](\Amc)$, then there exists a solution $\phi[0]$ to $\Hmc[0]$ such that~\eqref{eqn:phi0} holds.
In view of Assumption~\ref{ass:Amcs}, $\Amc[s]$ being strongly forward pre-invariant for $\Hmc$ implies that $\rge\phi\subset\Amc[s]$.
In view of Assumption~\ref{ass:Amc0} and~\cite[Theorem~7.12]{Goebel2012}, $\Amc[0]$ is also uniformly globally pre-asymptotically stable for $\Hmc[0]$. Then there exists $\alpha[0]\in\Kinf$ such that
\begin{align}\label{eqn:alpha_0}
\norm[{\Amc[0]}]{\phi[0](t,j)}\leq \alpha[0](\norm[{\Amc[0]}]{\phi[0](0,0)})
\end{align}
for each $(t,j)\in\dom\phi[0]$. The right-hand side of the inequality becomes zero when $\phi[0](0,0)\in\Amc[0]$, which implies that $\rge\phi[0]\subset\Amc[0]$. Notice that $\phi[0]^r$, a full $j$-reparametrization of $\phi[0]$, keeps its value unchanged when jumping through~\eqref{eqn:(x,x_c) by G_s}, namely $\rge\phi[0]^r=\rge\phi[0]$, and hence $\rge\phi[0]^r\subset\Amc[0]$.
Together with the relation $\rge\phi\subset\Amc[s]$, we obtain strong forward pre-invariance of $\Amc$ for $\Hmc$.

Next, we prove that each solution $\phi\in\Smc[\Hmc](\Amc+\mu\ball)$ is such that $(t,j)\mapsto\norm[\Amc]{\phi(t,j)}$ is bounded. Note that $\Amc+\mu\ball\subset(\Amc[s]+\mu\ball)\cap \ell\ball$ with $\ell>\sup_{\xi\in\Amc+\mu\ball}|\xi|$.
By Lemma~\ref{lem:phi}, there exists $(t^\star,j^\star)\in\dom\phi$ with $t^\star+j^\star\geq\settle(\phi(0,0))$ such that $(t,j)\mapsto \psi(t,j)\ceq\phi(t+t^\star,j+j^\star)$ for each $(t,j)\in\dom\psi\ceq\{(t,j)\in\Rnneg\times \naturals: (t+t^\star,j+j^\star)\in\dom\phi\}$ is complete and $\psi\in\Smc[\Hmc](\Amc[s])$.
Since solutions to a hybrid system are continuous by definition, and
\begin{align}\label{eqn:E}
\Emc\ceq\{(t,j)\in\dom\phi:t+j\leq t^\star+j^\star\}
\end{align}
is a compact hybrid time domain in view of the proof of Lemma~\ref{lem:phi}, then $\phi(\Emc)$ is compact and, in particular, bounded.
For each $(t,j)\in\dom\phi$ satisfying $t+j\geq t^\star+j^\star$, we have $\phi(t,j)\in\Amc[s]$ by strong forward pre-invariance of $\Amc[s]$ for $\Hmc$. Thus $\psi(t,j)\in\Amc[s]$ for each $(t,j)\in\dom\psi$.
By Lemma~\ref{lem:norm}, there exists a solution $\psi[0]$ to $\Hmc[0]$ such that
\begin{align}\label{eqn:norm after T}
\norm[\Amc]{\psi(t,j)}\leq \alpha(\norm[{\Amc[0]}]{\psi[0]^r(t,j)})
\end{align}
for each $(t,j)\in\dom\psi$.
Recall from Definition~\ref{def:jreparametrization} that
\begin{align}\label{eqn:rho_norm}
\norm[{\Amc[0]}]{\psi[0]^r(t,j)} = \norm[{\Amc[0]}]{\psi[0](t,\rho(j))}
\end{align}
for each $(t,j)\in\dom\psi$.
Since $(t,\rho(j))\in\dom\psi[0]$ and~\eqref{eqn:alpha_0} holds for each element in $\dom\psi[0]$, we have
\begin{align}\label{eqn:alpha_0_norm}
\norm[{\Amc[0]}]{\psi[0](t,\rho(j))} \leq \alpha[0](\norm[{\Amc[0]}]{\psi[0](0,0)})
\end{align}
for each $(t,j)\in\dom\psi$.
Combining~\eqref{eqn:norm after T},~\eqref{eqn:rho_norm}, and~\eqref{eqn:alpha_0_norm} results in
$\norm[\Amc]{\psi(t,j)} \leq \alpha\circ\alpha[0](\norm[{\Amc[0]}]{\psi[0](0,0)})$
for each $(t,j)\in\dom\psi$, where $\alpha\circ\alpha[0]\in\Kinf$.
Note that $\psi[0](0,0)$ is finite due to $\psi[0](0,0)=\psi[0]^r(0,0)=\bmtx{I&0}\phi(t^\star,j^\star)$, where $\phi(t^\star,j^\star)\in\phi(\Emc)$ with $\Emc$ defined by~\eqref{eqn:E}. Together with compactness of $\Amc$ and $\Amc[0]$, this shows boundedness of $\psi$, which in turn implies boundedness of $\phi$ since we have proved its boundedness at or before $(t^\star,j^\star)$. Hence we have boundedness of $(t,j)\mapsto\norm[\Amc]{\phi(t,j)}$.

We then prove that $\Amc$ is uniformly pre-attractive for $\Hmc$ from $\Amc+\mu\ball$. Since each $\phi\in\Smc[\Hmc](\Amc+\mu\ball)$ is complete in view of Lemma~\ref{lem:phi}, the hybrid system $\Hmc$ is said to be pre-forward complete (c.f.~\cite[Definition~6.12]{Goebel2012}). Since it satisfies the hybrid basic conditions by Assumption~\ref{ass:hbc}, it follows from compactness of $\Amc+\mu\ball$ and~\cite[Lemma~3.4]{Berk2020} that there exists a compact set $\Xmc\subset\Xi$ such that each solution $\phi\in\Smc[\Hmc](\Amc+\mu\ball)$ satisfies $\phi(t,j)\in \Xmc$ for each $(t,j)\in\dom\phi$ verifying
$$t+j\leq \settle_{\mu} \ceq\sup\settle(\Amc+\mu\ball)\leq\sup\settle((\Amc[s]+\mu\ball)\cap \ell\ball)$$
where $\settle_{\mu}$ is finite by Assumption~\ref{ass:Amcs}. We borrow from the previous paragraph the definitions of $\psi$ and $\psi[0]$. Observing that $\psi(0,0)\in\Xmc\cap\Amc[s]$, we consider the projection of $\Xmc\cap\Amc[s]$ onto $\reals[n_x+n_c]$ where the plant state and the controller state live in order to make use of solution properties of $\psi[0]$. To this end, define $\Xmc[0]\ceq\Pi(\Xmc\cap\Amc[s]|\reals[n_x+n_o])$,
which is compact.
Now for arbitrary $\varepsilon>0$, pick $\varepsilon_0\ceq\alpha\inv(\varepsilon)$ and $r_0\ceq\sup_{\xi[0]\in\Xmc[0]}\norm[{\Amc[0]}]{\xi[0]}$. It can be verified that $\varepsilon_0>0$ and $\norm[{\Amc[0]}]{\xi[0]}\leq r_0<+\infty$ for each $\xi[0]\in\Xmc[0]$. By uniform global pre-attractivity of $\Amc[0]$ for $\Hmc[0]$ in view of Assumption~\ref{ass:Amc0}, it follows that, for the chosen $\varepsilon_0$ there exists $\tau_0>0$ such that for each solution $\psi[0]$ to $\Hmc[0]$ from $\Xmc[0]$, $(t,j)\in\dom\psi[0]$ and $t+j\geq\tau_0$ imply $\norm[{\Amc[0]}]{\psi[0](t,j)}\leq\varepsilon_0$.
By~\eqref{eqn:rho_norm}, we have
\begin{align}\label{eqn:rho_norm_eps}
\norm[{\Amc[0]}]{\psi[0]^r(t,j)} = \norm[{\Amc[0]}]{\psi[0](t,\rho(j))}\leq\varepsilon_0
\end{align}
for each $(t,j)\in\dom\psi$ such that $t+\rho(j)\geq\tau_0$.
In the sequel, we show the existence of $\tau_1>0$ such that the above inequality holds for each $(t,j)\in\dom\psi$ such that $t+j\geq\tau_1$, where $\psi\in\Smc[\Hmc](\Xmc\cap\Amc[s])$ is arbitrary. Due to Assumption~\ref{ass:finite jumps}, the selection $r=\sup_{\xi\in\Xmc}|\xi|<+\infty$ leads to the fact that, there exist $\tau_{\Delta}\in\Rnneg$ and $N\in\naturals$ such that for each $\psi\in\Smc[\Hmc](\Xmc\cap\Amc[s])$, we have $\psi(0,0)\in r\ball$ and $j\leq\frac{t}{\tau_{\Delta}}+N$
for each $(t,j)\in\dom\psi$. Hence, for each $\psi\in\Smc[\Hmc](\Xmc\cap\Amc[s])$ and each $(t,j)\in\dom\psi$, $t+\rho(j)<\tau_0 \implies \tau_{\Delta}(j-N)+\rho(j)<\tau_0 \implies \tau_{\Delta}(j-N)<\tau_0 \implies j<\frac{\tau_0}{\tau_{\Delta}}+N$. This in turn implies that $t+j=t+\rho(j)+j-\rho(j)<\tau_0+j<\tau_0+\frac{\tau_0}{\tau_{\Delta}}+N =: \tau_1<+\infty$. Considering the contrapositive statement, we have $t+j\geq\tau_1$ implies $t+\rho(j)\geq\tau_0$ for each $(t,j)\in\dom\psi$ and each $\psi\in\Smc[\Hmc](\Xmc\cap\Amc[s])$, for which~\eqref{eqn:rho_norm_eps} holds.
For the chosen $\varepsilon$, select $\tau\ceq\settle_{\mu}+\tau_1$. Then for each $(t,j)\in\dom\phi$ such that $t+j\geq\tau$, we have $\norm[\Amc]{\phi(t,j)}\leq\alpha(\varepsilon_0)=\varepsilon$ by applying~\eqref{eqn:norm after T} to~\eqref{eqn:rho_norm_eps}. This proves uniform pre-attractivity of $\Amc$ for $\Hmc$ from $\Amc+\mu\ball$.

Local pre-asymptotic stability of $\Amc$ for $\Hmc$ follows from~\cite[Proposition~7.5]{Goebel2012} while uniformity comes from~\cite[Theorem~7.12]{Goebel2012}.
Finally, note from the proof of Lemma~\ref{lem:phi} that each $\phi\in\Smc[\Hmc]$ is complete, thereby concluding the proof.
\end{IEEEproof}

\begin{remark}
In contrast to our previous work where the assumption on the controller node also involves analysis of the sensor node (cf.~\cite[Assumption~3]{zhu_acc_2022}), here in this paper we make the stand-alone Assumption~\ref{ass:Amc0} for the controller. In other words, using the method in this paper, we gain an advantage in freely selecting the controller without having to consider its interconnection with the sensor.
However, note that there is a compromise between the separation in this work and the freedom to define $\Amc[s]$ in our previous work.

\end{remark}


In the following corollary, we demonstrate that the hybrid system $\Hmc$ does not have solutions with vanishing inter-event times and, in particular, is free of Zeno solutions.

\begin{corollary}\label{coro:Zeno}
If Assumptions~\ref{ass:Amc0}-\ref{ass:finite jumps} hold, then each $\phi\in\Smc[\Hmc](\Amc+\mu\ball)$ is not Zeno, where $\mu$ comes from Assumption~\ref{ass:Amcs}.
\end{corollary}
\begin{IEEEproof}
The hybrid system $\Hmc$ is nominally well-posed by Assumption~\ref{ass:hbc}. Meanwhile, each $\phi\in\Smc[\Hmc](\Amc+\mu\ball)$ is bounded as shown in the proof of Theorem~\ref{thm:stability}. Furthermore, there exists no complete discrete solution to $\Hmc$, otherwise Assumption~\ref{ass:finite jumps} implies that the existence of such a solution leads to a contradiction to Definition~\ref{def:finite jumps}.
Now, the desired conclusion follows from the contrapositive statement of \cite[Theorem~1]{Casau2022}.
\end{IEEEproof}

In the case where Assumption~\ref{ass:finite jumps} is relaxed by dropping the ``uniform semiglobal'' property, the following corollary gives a weaker stability result compared to Theorem~\ref{thm:stability}. {This result is particularly useful when it becomes hard or even impossible to prove the uniformity of the persistent flow time with respect to the initial states.}

\begin{assumption}\label{ass:finite jumps'}
Each solution to $\Hmc$ in~\eqref{eqn:closedloop} possess a persistent flow time from $\Amc[s]$ in~\eqref{eqn:Amc_s}.
\end{assumption}

\begin{corollary}\label{coro:attractiviy}
If Assumptions~\ref{ass:Amc0}-\ref{ass:Amcs} and \ref{ass:finite jumps'} hold, then $\Amc$ in~\eqref{eqn:Amc} is locally attractive for $\Hmc$ in~\eqref{eqn:closedloop}.
\end{corollary}
\begin{IEEEproof}
Consider $\mu$ from Assumption~\ref{ass:Amcs}. In view of the third and the last paragraphs of the proof of Theorem~\ref{thm:stability}, we have each $\phi\in\Smc[\Hmc](\Amc+\mu\ball)$ is bounded and complete. It remains to prove that $\lim_{t+j\to+\infty}\norm[\Amc]{\phi(t,j)}=0$. By Assumption~\ref{ass:finite jumps'}, for each $\phi\in\Smc[\Hmc](\Amc+\mu\ball)$, there exist $\tau_\Delta>0$ and $N\in\naturals$ satisfies $j\leq\frac{t}{\tau_\Delta}+N$ for each $(t,j)\in\dom\phi$. The rest of the proof follows similar arguments in the fourth paragraph of the proof of Theorem~\ref{thm:stability}, which renders the implication that for each $\varepsilon>0$ there exists $\tau>0$ such that $\norm[\Amc]{\phi(t,j)}\leq\varepsilon$ for each $(t,j)\in\dom\phi$ satisfying $t+j\geq\tau$. This concludes the proof.
\end{IEEEproof}

The following result provides a global version of Theorem~\ref{thm:stability}.
\begin{corollary}\label{coro:GAS}
If Assumptions~\ref{ass:Amc0},~\ref{ass:hbc}, and \ref{lem:norm} hold, then $\Amc$ in~\eqref{eqn:Amc} is globally uniformly globally asymptotically stable for $\Hmc$ in~\eqref{eqn:closedloop} if Assumption~\ref{ass:Amcs} holds for each $\mu>0$.
\end{corollary}
\begin{IEEEproof}
Stability of $\Amc$ for $\Hmc$ comes from Theorem~\ref{thm:stability} with an arbitrary $\mu>0$. From the fourth paragraph of the proof of Theorem~\ref{thm:stability}, we have $\Amc$ is uniformly attractive for $\Hmc$ from $\Amc+\mu\ball$ for each $\mu>0$, implying uniform global attractivity of $\Amc$ for $\Hmc$. Global asymptotic stability of $\Amc$ for $\Hmc$ then follows, which implies that $\Amc$ is uniformly globally asymptotically stable for $\Hmc$ in view of~\cite[Theorem~7.12]{Goebel2012}.
\end{IEEEproof}

Based on Corollary~\ref{coro:GAS}, we demonstrate that $\Hmc$ is endowed with intrinsic robustness which stems from meeting the hybrid basic conditions. In this direction, we consider a perturbed hybrid system $\Hmc[\rho]\ceq(\Cmc[\rho],\Fmc[\rho],\Dmc[\rho],\Gmc[\rho])$ in the form of \cite[Definition~6.27]{Goebel2012} defined by
\begin{equation}\label{eqn:H rho}
\begin{aligned}
\Fmc[\rho](\xi)&\ceq \con\Fmc((\xi+\rho\ball)\cap\Cmc) + \rho\ball&&\forall\xi\in \Cmc[\rho]\\
\Gmc[\rho](\xi)&\ceq \Gmc((\xi+\rho\ball)\cap\Dmc)+\rho\ball&&\forall\xi\in \Dmc[\rho]
\end{aligned}
\end{equation}
where $\Cmc[\rho]\ceq \{\xi\in\Xi:(\xi+\rho\ball)\cap\Cmc\neq\emptyset\}$, $\Dmc[\rho]\ceq \{\xi\in\Xi:(\xi+\rho\ball)\cap\Dmc\neq\emptyset\}$,
$\rho>0$, and $\con$ denotes the closure of convex hull.
\begin{corollary}\label{coro:robust}
If Assumptions~\ref{ass:Amc0},~\ref{ass:hbc}, and \ref{lem:norm} hold while Assumption~\ref{ass:Amcs} holds for each $\mu>0$, then there exists $\beta\in\KL$ such that for each compact set $\Xmc\subset\Xi$ and every $\varepsilon>0$, there exists $\rho>0$ such that every $\phi\in\Smc[{\Hmc[\rho]}](\Xmc)$ satisfies
\begin{equation}\label{eqn:robust}
\norm[{\Amc}]{\phi(t,j)} \leq \beta(\norm[{\Amc}]{\phi(0,0)},t+j)+\varepsilon\quad\forall (t,j)\in\dom\phi
\end{equation}
\end{corollary}
\begin{IEEEproof}
By Assumption~\ref{ass:hbc}, $\Hmc$ is nominally well-posed. Since the compact set $\Amc$ is globally asymptotically stable by Corollary~\ref{coro:GAS}, it follows from \cite[Theorem~7.12]{Goebel2012} that $\Amc$ is $\KL$ asymptotically stable on $\Xi$. Hence, $\Amc$ is semiglobally practically robustly $\KL$ asymptotically stable on $\Xi$ in view of \cite[Lemma~7.20]{Goebel2012}. In particular, the specific choice $\omega:\Xi\to\Rnneg$ defined by $\omega(x)=\norm[{\Amc}]{x}$ in \cite[Definition~7.18]{Goebel2012} allows one to arrive at the inequality~\eqref{eqn:robust} for the perturbed hybrid system $\Hmc[\rho]$.
\end{IEEEproof}
	{Although the intrinsic robustness property is reflected by a \emph{non-constructive} proof of the positive margin $\rho$, it is explicitly stated in~\cite{Goebel2012} as follows: \emph{``at times it might be possible to guarantee that the actual property for the nominal system still holds, at least practically, when the disturbances are small.''} This means we can readily model measurement noises, state perturbations, model uncertainties with a maximum magnitude $\rho$, similar to the discussions in~\cite{Mayhew2011}.}

{\section{Application}\label{sec:application}
We illustrate the versatility of our method through implementing model-based ETC on two existing but different hybrid controllers, resulting in novel control schemes. The following numerical results are obtained from MATLAB/Simulink software with the hybrid equation solver~\cite{sanfelice_toolbox_2013}.
\subsection{Implementation of a sampled-data controller}\label{sec:application linear}}
We demonstrate Theorem~\ref{thm:stability}, Corollary~\ref{coro:GAS}, and Corollary~\ref{coro:robust} through the application example in~\cite{zhu_acc_2022}, where model-based ETC is implemented on a sampled-data controller (as a hybrid controller). We extend the results therein from a specific linear plant to general linear plants.
For sake of brevity, we do not present the dynamics of each system component here. Rather, we adopt the results in this paper to construct a simpler stability proof compared to~\cite[Propositions~1-3]{zhu_acc_2022}.
Specifically, we have the following result.
\begin{corollary}[{Extension of~\cite[Section~V]{zhu_acc_2022}}]\label{coro:app1}
	  Given a stabilizable pair $(A,B)$ in~\cite[Eqn.~(7)]{zhu_acc_2022}, there exists $K$ in~\cite[Eqn.~(8)]{zhu_acc_2022} such that $\Amc$ is globally asymptotically stable for $\Hmc$ with intrinsic robustness.
\end{corollary}
\begin{IEEEproof}
	Due to~\cite[Assumption~4]{zhu_acc_2022}, it follows from~\cite[Theorem~3]{Postoyan2015} that Assumption~\ref{ass:Amc0} holds.
	By virtue of~\cite[Proposition~1]{zhu_acc_2022}, Assumption~\ref{ass:hbc} holds for $\Hmc$.
	\cite[Proposition~2]{zhu_acc_2022} implies that Assumption~\ref{ass:Amcs} holds.
	By construction, there are no jumps due to satisfaction of the event-triggering condition for each solution with initial state in $\Amc[s]$ defined similarly to~\cite[Eqn.~(11)]{zhu_acc_2022}. On one hand, note that both the sampled-data controller and the hybrid observer issue periodic jumps. On the other hand, note that there are no more than four consecutive jumps occurring at the same continuous time. Therefore, there exist $\tau>0$ and $N=4\in\naturals$ such that $j\leq\frac{t}{\tau}+N$ in Definition~\ref{def:finite jumps} holds, implying that Assumption~\ref{ass:finite jumps} holds.
	Now applying Theorem~\ref{thm:stability} and Corollary~\ref{coro:GAS} concludes the proof on the stability part while applying Corollary~\ref{coro:robust} concludes the robustness part.
\end{IEEEproof}

By replacing~\cite[Assumption~3]{zhu_acc_2022} with Assumption~\ref{ass:Amc0} and Assumption~\ref{ass:finite jumps}, we separate the controller design from the sensor design, which is the major improvement to~\cite{zhu_acc_2022} as mentioned in Remark~\ref{rem:Amc0}. 

\begin{figure}[ht!]
	\centering
	\input{app1.tex}
	\includegraphics[width=0.7\textwidth]{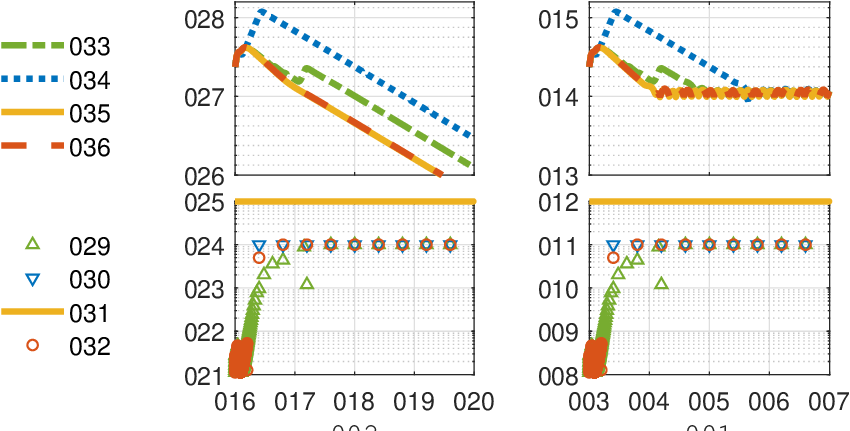}
	\caption{Comparison between continuous-time evolution and the corresponding inter-transmission time of solutions to the unperturbed/perturbed closed-loop systems using our proposed controller and the controllers (a)-(c) in Section~\ref{sec:application linear}. Vertical axes are in log scale.
	Left column of figures: plots for the unperturbed closed-loop systems.
	Right column of figures: plots for the perturbed closed-loop systems.
	Top row of figures: continuous-time evolution of the distance of solutions to $\Amc$ in Corollary~\ref{coro:app1}.
	Bottom row of figures: inter-transmission time for the sensor-to-controller communication channels.}
	\label{fig:app1}
\end{figure}

We demonstrate Corollary~\ref{coro:app1} through Fig.~\ref{fig:app1} with an unperturbed/perturbed linearized model of a batch reactor, which is a benchmark example in the networked control system literature (cf.~\cite{Heemels2013}). The process of obtaining the parameters for the sampled-data controller and the rest parameters follow similar lines in~\cite[Section~VI]{zhu_acc_2022} and~\cite[Section~8]{Zhu2024}, respectively.
The perturbed linearized model of the batch reactor is modeled by considering in~\eqref{eqn:H rho} the state perturbation $d:\dom\xi\to\Xi$ defined by $d(t,j)=\rho\sin(t)(1,1,\dots,1)$ with $\rho=10^{-3}$.
The control performance of our model-based event-triggered controller is compared with: (a) one with model-based event-triggered transmission with asymptotic convergent observers (cf.~\cite{Heemels2013}); (b) one with time-triggered transmission (cf.~\cite{Borri2024}) whose periodicity equals the limit inferior of the inter-transmission time of our proposed controller; (c) and one with continuous transmission.
The left/right column of Fig.~\ref{fig:app1} shows the continuous-time evolution of solutions as well as the inter-transmission time for the unperturbed/perturbed closed-loop systems using the aforementioned controllers. We observe that in both cases (unperturbed/perturbed), our proposed controller outperforms the controllers (a) and (b) in terms of convergence performance and achieve similar convergence pattern as the controller (c), while not issuing, on average, more transmissions than any of the controllers (a)-(c).

{\subsection{Implementation of a synergistic controller}\label{sec:application nonlinear}}
We demonstrate Corollary~\ref{coro:attractiviy} through the problem in~\cite{Gui2021} on quaternion-based attitude control of a rigid body without velocity measurements, where a hybrid finite-time convergent observer recovers the velocity to feed a synergistic controller. {We extend their results to consider the existence of a sensor-to-controller communication channel, which is handled by implementing our proposed model-based ETC on the synergistic controller (as a hybrid controller).}
Specifically, the plant with state $x\ceq(\mf{q},\omega)\in\sphere{3}\x\reals[3]\ceqinv\Xmc$ is defined by
\begin{align}\label{eqn:quat}
\dot{\mf{q}}=E(\mf{q})\omega,\quad
\dot{\omega}=u,\quad
y=\mf{q}
\end{align}
where $\mf{q}\ceq(\mf{n},\mf{e})$ denotes the attitude, $\omega$ denotes the angular velocity, $u$ is the input, and $y$ is the output. The function $E:\sphere{3}\to\reals[4\x 3]$ is defined by $E(\mf{q})\ceq\frac{1}{2}(-\mf{e}\tp,\mf{n}I+\sk(\mf{e}))$
for each $\mf{q}\in\sphere{3}$, where $\sk:\reals[3]\to\so{3}\ceq\{M\in\reals[3\x 3]:M=-M\tp\}$ verifies $S(v)w=v\x w$ for each $v,w\in\reals[3]$ with the cross product operator $\x$. 
The synergistic controller with state $x_c\in\Xmc[c]\ceq\{-1,1\}$ is defined by
\begin{equation}\label{eqn:quat_con}
\begin{aligned}
\dot{x}_c&= 0 \,\,\,\,\,\,\,\,\, (x_c,x)\in \Cmc[c]\ceq\{(x_c,x)\in\Xmc[c]\x\Xmc:x_c\mf{n}\geq-\delta[c]\}\\
x_c\pl&= -x_c \, (x_c,x)\in \Dmc[c]\ceq\{(x_c,x)\in\Xmc[c]\x\Xmc:x_c\mf{n}\leq-\delta[c]\}
\end{aligned}
\end{equation}
with some $\delta[c]\in(0,1)$. The hybrid controller outputs the control input $u$ through the feedback law $\kappa:\Xmc[c]\x\Xmc\to\reals[3]$ defined by
\begin{equation}\label{eqn:quat_u}
    u=\kappa(x_c,x)\ceq-\varphi_{\beta[1]}(x_c\mf{q})-\signpower{\omega}{\beta[2]}
\end{equation}
for each $(x_c,x)\in\Xmc[c]\x\Xmc$, where $\beta[1]\in(0,1)$, $\beta[2]\ceq\frac{2-2\beta[1]}{2-\beta[1]}$, the continuous function $\varphi_{\beta}:\sphere{3}\to\reals[3]$ for a given $\beta\in(0,1)$ is defined by $ \varphi_{\beta}(\mf{q})=\mf{e}(2-2\mf{n})^{-\beta/2}$ for each $\mf{n}\in[0,1)$ and $ \varphi_{\beta}(\mf{q})=0$ for each $\mf{n}=1$ with
$\mf{q}\ceq(\mf{n},\mf{e})\in\sphere{3}$, and {the continuous function
$\signpower{\cdot}{\beta}:\reals[n]\to\reals[n]$ for a given $\beta\in(0,1)$ is defined by 
$\signpower{v}{\beta}=(\mathrm{sign}(v_1)|v_1|^{\beta},\mathrm{sign}(v_2)|v_2|^{\beta},\cdots,\mathrm{sign}(v_n)|v_n|^{\beta})$
for each $v\in\reals[n]$.}
The hybrid finite-time convergent observer with state $\xo\ceq(\hat{\mf{q}},\hat{\omega},\tau)\in\Xmc\x\reals\ceqinv\Xmc[o]$ is defined by
\begin{equation}\label{eqn:quat_obs}
\begin{aligned}
&\left.\begin{aligned}
\dot{\hat{\mf{q}}}&=E(\hat{\mf{q}})(\hat{\omega}+\varphi_{\alpha[1]}(\hat{\mf{q}}^\ast\otimes y))\\
\dot{\hat{\omega}}&=u+\varphi_{\alpha[2]}(\hat{\mf{q}}^\ast\otimes y)\\
\dot{\tau}&=0
\end{aligned}\right\rbrace (\xo,y,u)\in\Cmc[o,1]\cap\Cmc[o,2]\\
&
\xo\pl\in\Gmc[o,1](\xo)\cup\Gmc[o,2](\xo)\qquad\,\,\,\,(\xo,y,u)\in\Dmc[o,1]\cup\Dmc[o,2]
\end{aligned}
\end{equation}
where $\alpha[1]\in(0,\frac{1}{2})$, $\alpha[2]\ceq 2\alpha[1]$,
\begin{align}
    \Cmc[o,1]&\ceq\{(\xo,y,u)\in\Xmc[o]\x\Xmc:\mf{1}\tp(\hat{\mf{q}}^\ast\otimes y)\geq-\delta[o]\}\\
    \Cmc[o,2]&\ceq\{(\xo,y,u)\in\Xmc[o]\x\Xmc:\tau\in[0,\bar{\tau}]\}\\
    \Dmc[o,1]&\ceq\{(\xo,y,u)\in\Xmc[o]\x\Xmc:\mf{1}\tp(\hat{\mf{q}}^\ast\otimes y)\leq-\delta[o]\}\\
    \Dmc[o,2]&\ceq\{(\xo,y,u)\in\Xmc[o]\x\Xmc:\tau=\bar{\tau}\}
\end{align}
with $\delta[o]\in(0,1)$ and $\bar{\tau}>0$, and the maps
\begin{align}
    \Gmc[o,1](\xo)&\ceq
        (-\hat{\mf{q}},\hat{\omega},\tau)&&\forall\xo\in\Pi(\Dmc[o,1]|\Xmc)\\
    \Gmc[o,2](\xo)&\ceq(\hat{\mf{q}},\hat{\omega},0)&&\forall\xo\in\Pi(\Dmc[o,2]|\Xmc)
\end{align}
with $\dom\Gmc[o,1]=\Pi(\Dmc[o,1]|\Xmc)$ and $\dom\Gmc[o,2]=\Pi(\Dmc[o,2]|\Xmc)$, the operator $^\ast$ denotes the conjugation of a quaternion, namely $\mf{q}^\ast=(\mf{n},-\mf{e})$ for a given $\mf{q}=(\mf{n},\mf{e})$, and the operator $\otimes$ denotes the multiplication of two quaternions, namely $\mf{q}_1\otimes\mf{q}_2\ceq(\mf{n}_1\mf{n}_2-\mf{e}_1\tp\mf{e}_2,\mf{n}_1\mf{e}_2+\mf{n}_2\mf{e}_1+\sk(\mf{e}_1)\mf{e}_2)$
for each $\mf{q}_1,\mf{q}_2\in\sphere{3}$, where the multiplicative identity is denoted as $\mf{1}\ceq(1,0,0,0)$. The observer outputs $\hat{x}$ through the law $h_o:\Xmc[o]\x\sphere{3}\to\Xmc[o]$ defined by $\hat{x}=h_o(\xo,y)\ceq(\mf{q},\hat{\omega},\tau)$
for each $(\xo,y)\ceq(\hat{\mf{q}},\hat{\omega},\tau,\mf{q})\in\Xmc[o]\x\sphere{3}$.
The synthetic model of the plant with state $\xs\ceq(\mf{q}_s,\omega[s])\in\Xmc$ is defined by
\begin{equation}\label{eqn:quat_syn}
\begin{aligned}
&\begin{aligned}
\dot{x}_s&=(E(\mf{q}_s)\omega[s],u)
\end{aligned}&&(\xs,u,\hat{x})\in\Cmc[s,1]\cap\Cmc[s,2]\\
&\begin{aligned}
\xs\pl&=(y,\hat{\omega})
\end{aligned}&&(\xs,u,\hat{x})\in\Dmc[s,1]\cup\Dmc[s,2]
\end{aligned}
\end{equation}
where
\begin{align}
    \Cmc[s,1]&\ceq\{(\xs,u,\hat{x})\in\Xmc\x\reals[3]\x\Xmc[o]:|\omega[s]-\hat{\omega}|\leq\delta\}\\
    \Cmc[s,2]&\ceq\{(\xs,u,\hat{x})\in\Xmc\x\reals[3]\x\Xmc[o]:\tau\in[0,\bar{\tau}]\}\\
    \Dmc[s,1]&\ceq\{(\xs,u,\hat{x})\in\Xmc\x\reals[3]\x\Xmc[o]:|\omega[s]-\hat{\omega}|\geq\delta\}\\
    \Dmc[s,2]&\ceq\{(\xs,u,\hat{x})\in\Xmc\x\reals[3]\x\Xmc[o]:\tau=\bar{\tau}\}
\end{align}
with $\delta>0$, which means that the event-triggering condition issues a transmission either when the norm of the synthetic angular velocity deviates too much from that of the estimated angular velocity, or when the timer $\tau$ reaches its periodicity.
Following the proposed scheme in Section~\ref{sec:mbetc}, we obtain the hybrid system with state $(x,\xc)\in\Xmc\x\Xmc[c]$ resulting from the interconnection of the plant~\eqref{eqn:quat} without output and the synergistic controller~\eqref{eqn:quat_con} through the feedback law~\eqref{eqn:quat_u}, denoted by $\Hmc[0]\ceq(\Cmc[0],\Fmc[0],\Dmc[0],\Gmc[0])$ and defined by
\begin{equation}\label{eqn:quat_H0}
\begin{aligned}
    \Fmc[0](x,\xc)&\ceq(E(\mf{q})\omega,\kappa(\xc,x),0) & &\forall (x,\xc)\in\Cmc[0]\\
    \Gmc[0](x,\xc)&\ceq(x,-\xc) & &\forall (x,\xc)\in\Dmc[0]
\end{aligned}
\end{equation}
where $\Cmc[0]\ceq\{(x,\xc)\in\Xmc\x\Xmc[c]:(\xc,x)\in\Cmc[c]\}$ and $\Dmc[0]\ceq\{(x,\xc)\in\Xmc\x\Xmc[c]:(\xc,x)\in\Dmc[c]\}$.
Also, we obtain the hybrid system with state $\xi\ceq(x,\xc,\xs,\xo)\in\Xmc\x\Xmc[c]\x\Xmc\x\Xmc[o]\ceqinv\Xi$ resulting from the interconnection of the plant~\eqref{eqn:quat}, the hybrid controller~\eqref{eqn:quat_con} with the feedback law~\eqref{eqn:quat_u}, the hybrid observer~\eqref{eqn:quat_obs}, and the synthetic model~\eqref{eqn:quat_syn}, denoted by $\Hmc\ceq(\Cmc,\Fmc,\Dmc,\Gmc)$ and defined by
\begin{equation}
\begin{aligned}\label{eqn:quat_H}
\Fmc(\xi)&\ceq\pmtx{(E(\mf{q})\omega,\kappa(\xc,\xs))\\0\\(E(\mf{q}_s)\omega[s],\kappa(\xc,\xs))\\E(\hat{\mf{q}})(\hat{\omega}+\varphi_{\alpha[1]}(\hat{\mf{q}}^\ast\otimes \mf{q}))\\\kappa(\xc,\xs)+\varphi_{\alpha[2]}(\hat{\mf{q}}^\ast\otimes \mf{q})} & &\forall \xi\in\Cmc\\
\Gmc(\xi)&\ceq\Gmc[c](\xi)\cup \bigcup_{i\in\{1,2,3\}}\Gmc[so,i](\xi) & &\forall \xi\in\Dmc
\end{aligned}
\end{equation}
where $\Cmc\ceq\{\xi\in\Xi:(\xo,\mf{q},\kappa(\xc,\xs))\in\Cmc[o,1]\cap\Cmc[o,2], (\xc,\xs)\in\Cmc[c],(\xs,\kappa(\xc,\xs), h_o(\xo,\mf{q}))\in\Cmc[s,1]\cap\Cmc[s,2]\}$, $\Dmc\ceq\{\xi\in\Xi:(\xc,\xs)\in\Dmc[c]\text{ or }\xi\in\bigcup_{i\in\{1,2,3\}}\Dmc[so,i]\}$ with
\begin{align*}
    \Dmc[so,1]&\ceq\{\xi\in\Xi:(\xo,\mf{q},\kappa(\xc,\xs))\in\Dmc[o,1]\}\\
    \Dmc[so,2]&\ceq\{\xi\in\Xi:(\xs,\kappa(\xc,\xs),h_o(\xo,\mf{q}))\in\Dmc[s,1]\}\\
    \Dmc[so,3]&\ceq\{\xi\in\Xi:\tau=\bar{\tau}\}
\end{align*}
and $\Gmc[c],\Gmc[so,i]:\Xi\tto\Xi$ are defined by
\begin{align*}
    \Gmc[c](\xi)&\ceq(x,-\xc,\xs,\xo)&&\forall\xi\in\Xmc\x\Dmc[c]\x\Xmc[o]\\
    \Gmc[so,1](\xi)&\ceq(x,\xc,\mf{q}_s,\omega[s],-\hat{\mf{q}},\hat{\omega},\tau)&&\forall \xi\in\Dmc[so,1]\\
    \Gmc[so,2](\xi)&\ceq(x,\xc,\mf{q},\hat{\omega},\hat{\mf{q}},\hat{\omega},\tau)&&\forall \xi\in\Dmc[so,2]\\
    \Gmc[so,3](\xi)&\ceq(x,\xc,\mf{q},\hat{\omega},\hat{\mf{q}},\hat{\omega},0)&&\forall \xi\in\Dmc[so,3]
\end{align*}
with $\dom\Gmc[c]=\Xmc\x\Dmc[c]\x\Xmc[o]$ and $\dom\Gmc[so,i]=\Dmc[so,i]$ for each $i\in\{1,2,3\}$.

Before stating the results, we define the following sets of interest.
	\begin{equation}\label{eqn:quat_A0}
	\Amc[0]\ceq\{(x,\xc)\in\Xmc\x\Xmc[c]:x=(x_c\mf{1},0)\}
	\end{equation}
	\begin{equation}\label{eqn:quat_As}
	\Amc[s]\ceq\{\xi\in\Xi:x=\xs,\xo\in\{x\}\x[0,\bar{\tau}]\}
	\end{equation}
	\begin{equation}\label{eqn:quat_A}
	\Amc\ceq\{\xi\in\Xi:x=\xs=(\hat{\mf{q}},\hat{\omega})=(x_c\mf{1},0),\tau\in[0,\bar{\tau}]\}
	\end{equation}
\begin{corollary}[{Extension of~\cite{Gui2021}}]\label{coro:app2}
	$\Amc$ in~\eqref{eqn:quat_A} is globally attractive for $\Hmc$ in~\eqref{eqn:quat_H}.
\end{corollary}
\begin{IEEEproof}
	Due to~\cite[Theorem~9]{Gui2021}, it follows from~\cite[Remark~3.7]{Li2019} that Assumption~\ref{ass:Amc0} holds for $\Amc[0]$ in~\eqref{eqn:quat_A0} and $\Hmc[0]$ in~\eqref{eqn:quat_H0}.
	By construction, Assumption~\ref{ass:hbc} holds for $\Hmc$.
	It follows from~\cite{Gui2021} that there exists a settling-time function $\settle:\Xi\to[0,+\infty)$ satisfying $\sup\settle((\Amc[s]+\mu\ball)\cap\ell\ball)<+\infty$ for each $\mu,\ell>0$, implying that Assumption~\ref{ass:Amcs} holds for $\Amc[s]$ in~\eqref{eqn:quat_As} and $\Hmc$.
	By construction, there are no jumps by $\Gmc[so,1]$ nor $\Gmc[so,2]$ for each solution with initial state in $\Amc[s]$.
	On one hand, note that there are finite number of jumps by $\Gmc[c]$ and the jumps by $\Gmc[so,3]$ are evenly separated in continuous time by $\bar{\tau}$.
	On the other hand, note that there are no more than two consecutive jumps occurring at the same continuous time.
	Therefore, there exist $\tau>0$ and $N=4\in\naturals$ such that $j\leq\frac{t}{\tau}+N$ in Definition~\ref{def:finite jumps} holds, implying that Assumption~\ref{ass:finite jumps'} holds.
	Now applying Corollary~\ref{coro:attractiviy} concludes the proof.
\end{IEEEproof}

By implementing model-based ETC on a synergistic controller, we eliminate the need for continuous transmission on the sensor-to-controller communication channel in~\cite{Gui2021}, which marks the major improvement in terms of practical implementability.

\begin{figure}[ht!]
	\centering
	\input{app2.tex}
	\includegraphics[width=0.7\textwidth]{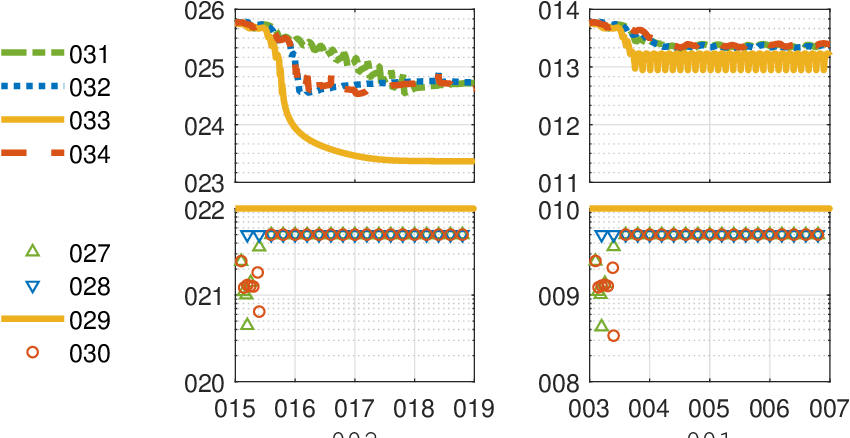}
	\caption{Comparison between continuous-time evolution and the corresponding inter-transmission time of solutions to the unperturbed/perturbed closed-loop systems using our proposed controller and the controllers (a)-(c) in Section~\ref{sec:application nonlinear}. Vertical axes are in log scale.
	Left column of figures: plots for the unperturbed closed-loop systems.
	Right column of figures: plots for the perturbed closed-loop systems.
	Top row of figures: continuous-time evolution of the distance of solutions to $\Amc$ in Corollary~\ref{coro:app2}.
	Bottom row of figures: inter-transmission time for the sensor-to-controller communication channels.}
	\label{fig:app2}
\end{figure}

We demonstrate Corollary~\ref{coro:app2} through Fig.~\ref{fig:app2} with the unperturbed system~\eqref{eqn:quat_H} and its perturbed one.
The parameters $\alpha[1]=\beta[1]=\frac{1}{3}$, $\delta[c]=\delta[o]=\frac{1}{2}$, $\delta=1$, and $\bar{\tau}=5$.
The perturbed system of~\eqref{eqn:quat_H} is modeled by considering in~\eqref{eqn:H rho} the state perturbation $d:\dom\xi\to\Xi$ defined by $d(t,j)=\rho\sin(t)(1,1,\dots,1)$ with $\rho=10^{-3}$.
The control performance of our model-based event-triggered controller is compared with: (a) one with model-based event-triggered transmission with asymptotic convergent observers (cf.~\cite{Dong2019}); (b) one with time-triggered transmission (cf.~\cite{Borri2024}) whose periodicity equals the limit inferior of the inter-transmission time of our proposed controller; (c) and one with continuous transmission (cf.~\cite{Gui2021}).
The left/right column of Fig.~\ref{fig:app2} shows the continuous-time evolution of solutions as well as the inter-transmission time for the unperturbed/perturbed closed-loop systems using the aforementioned controllers. We observe that in both cases (unperturbed/perturbed), our proposed controller outperforms the controllers (a) and (b) in terms of convergence performance, while not issuing, on average, more transmissions than any of the controllers (a)-(c).

{Although intrinsic robustness cannot be directly inferred from Corollary~\ref{coro:app2}, note that in Fig.~\ref{fig:app2} the proposed control scheme shows some robustness against the state perturbation $d$, which demonstrates certain practicality and implementation feasibility in applying our method to real-world control systems like a satellite in~\cite{Gui2021}.
}

\section{Conclusion}\label{sec:conclusion}
Through the lens of hybrid system tools, we show that the class of finite-time observer-based model-based event-triggered controllers in this paper renders a compact set asymptotically stable for the hybrid closed-loop system, which we use to solve two practical problems by implementing the proposed strategy based on some known stabilizing hybrid controllers.
The major advantages of our method are that: 1) it offers some degree of separation between the controller design and the sensor design; and 2) it allows the consideration of a wide class of general nonlinear plants, hybrid controllers, hybrid sensors, and model-based event-triggering conditions.
The limitations of our method are that: 1) it assumes synchronization between the copies of the controller and of the synthetic model; 2) it does not guarantee the existence of a positive lower bound of the inter-transmission time; and 3) it does not achieve a full separation principle between the controller design and the sensor design.
Some possible applications of our method can be: 1) extension to decentralized and/or distributed control settings; 2) investigation of combining various hybrid controllers, various finite-time convergent hybrid observers, and various event-triggering conditions; 3) and adaption to practical problems for which transmission and/or computation cost is a critical consideration, such as regulation tasks for smart grids, cooperative tasks for robots, information exchange among communication systems, to name a few.

\bibliographystyle{ieeetr}
\bibliography{biblio}

\vspace{-25 mm}

\end{document}